\documentclass[fleqn,usenatbib]{mnras}

\usepackage[T1]{fontenc}
\DeclareRobustCommand{\VAN}[3]{#2}
\let\VANthebibliography\thebibliography
\def\thebibliography{\DeclareRobustCommand{\VAN}[3]{##3}\VANthebibliography}

\newcommand{\sn}{\ifmmode {\rm S/N}\else${\rm S/N}$\fi}
\newcommand{\Pm}{\ifmmode P_{\rm m}\else$P_{\rm m}$\fi}
\newcommand{\Pf}{\ifmmode P_{\rm F}\else$P_{\rm F}$\fi}
\newcommand{\Po}{\ifmmode P_{\rm 1O}\else$P_{\rm 1O}$\fi}
\newcommand{\Poo}{\ifmmode P_{\rm 2O}\else$P_{\rm 2O}$\fi}
\newcommand{\Px}{\ifmmode P_{\rm x}\else$P_{\rm x}$\fi}
\newcommand{\Psh}{\ifmmode P_{\rm sh}\else$P_{\rm sh}$\fi}
\newcommand{\fF}{\ifmmode \nu_{\rm F}\else$\nu_{\rm F}$\fi}
\newcommand{\fo}{\ifmmode \nu_{\rm 1O}\else$\nu_{\rm 1O}$\fi}
\newcommand{\fx}{\ifmmode \nu_{\rm x}\else$\nu_{\rm x}$\fi}
\newcommand{\fsh}{\ifmmode \nu_{\rm sh}\else$\nu_{\rm sh}$\fi}
\renewcommand{\fm}{\ifmmode \nu_{\rm m}\else$\nu_{\rm m}$\fi}
\newcommand{\Af}{\ifmmode A_{\rm F}\else$A_{\rm F}$\fi}
\newcommand{\Ao}{\ifmmode A_{\rm 1O}\else$A_{\rm 1O}$\fi}
\newcommand{\Aoo}{\ifmmode A_{\rm 2O}\else$A_{\rm 2O}$\fi}
\newcommand{\Ax}{\ifmmode A_{\rm x}\else$A_{\rm x}$\fi}
\newcommand\redsout{\bgroup\markoverwith{\textcolor{red}{\rule[0.5ex]{2pt}{0.4pt}}}\ULon}
\usepackage{graphicx}	
\usepackage{amsmath}	
\usepackage{amssymb}	

\usepackage{newtxmath}
\usepackage{ulem}
\usepackage[dvipsnames]{xcolor}

\title[Frequency analysis of Cepheids in Galactic fields]{Frequency analysis of OGLE-IV photometry for classical Cepheids in Galactic fields: non-radial modes and modulations}

\author[R.S. Rathour, R. Smolec and H. Netzel]{
Rajeev Singh Rathour,$^{1}$\thanks{E-mail: rajeevsr@camk.edu.pl}
Rados\l{}aw Smolec$^{1}$, Henryka Netzel$^{1}$
\\
$^{1}$Nicolaus Copernicus Astronomical Centre, Polish Academy of Sciences, Bartycka 18, 00-716 Warszawa, Poland
}

\date{Accepted XXX. Received YYY; in original form ZZZ}

\pubyear{2020}

\begin{document}
\label{firstpage}
\pagerange{\pageref{firstpage}--\pageref{lastpage}}
\maketitle

\begin{abstract}
We analyse photometry of $\sim$2000 Galactic Cepheids available in the OGLE Collection of Variable Stars. We analyse both Galactic disk and Galactic bulge fields; stars classified both as single- and multi-periodic. Our goal was to search for additional low-amplitude variability. 

We extend the sample of multi-mode radial pulsators by identifying ten new candidates for double-mode and six new candidates for triple-mode pulsation. In the first overtone OGLE sample, we found twelve Cepheids with additional periodicity having period ratio $\Px/\Po\in (0.60,\, 0.65)$. These periodicities do not correspond to any other radial mode. While such variables are abundant in the Magellanic Clouds, only one Cepheid of this class was known in the Galaxy before our analysis. Comparing our sample with the Magellanic Cloud Cepheids we note a systematic shift towards longer pulsation periods for more metal rich Galactic stars. Moreover in eleven stars we find one more type of additional variability, with characteristic frequencies close to half of that reported in the group with (0.60,\, 0.65) period ratios. Two out of the above inventory show simultaneous presence of both signals.  Most likely, origin of these signals is connected to excitation of non-radial pulsation modes.

We report three Cepheids with low-amplitude periodic modulation of pulsation: two stars are single-mode fundamental and first overtone Cepheids and one is a double-mode Cepheid pulsating simultaneously in fundamental and in first overtone modes. Only the former mode is modulated. It is a first detection of periodic modulation of pulsation in this type of double-mode Cepheids.
\end{abstract}




\section{INTRODUCTION} 
\label{sec: Introduction}

Cepheids are intermediate to high-mass stars  ($\sim$3--13\,M$_{\odot}$) which undergo core helium-burning phase. They are amongst the most important class of intrinsic variable stars in the local universe. Cepheids exhibit pulsations of period roughly in the range 1--100 days and obey well-defined Period-Luminosity ($P-L$) relationship \citep{leavitt19081777}. They are ubiquitous and bright which makes them easily observable systems to study. Moreover, they have stable light curves over the subsequent pulsation cycle. Over the years, Cepheids have been found to be pulsating in radial modes: fundamental mode (F), first overtone (1O), double-mode (such as F+1O, 1O+2O) and even triple-mode (F+1O+2O, 1O+2O+3O) \citep[eg.,][]{moskalik2004nonradial,soszynski2008optical,soszynski2010optical,soszynski2011optical,soszynski2016multi}. Their importance extends from extracting knowledge about the conditions at the stellar interior and deriving physical parameters \citep[such as mass, radius, temperature and metallicity;][]{MoskalikDziembowski2005A&A...434.1077M,Beaulieu2006ApJ...653L.101B,Pilecki2018ApJ...862...43P,DeSomma2020ApJS..247...30D} to even cosmological probes for measuring the Hubble constant \citep{freedman2001final, freedman2012carnegie, riess20113, riess20162, riess2019large} using the $P-L$ relationship. \\

The modes that Cepheid pulsate in, were earlier thought to be only radial ones but this notion was amended by a study of Magellanic Clouds, in particular the Large Magellanic Cloud (LMC), eg., \cite{Moskalik&Kolaczkowski2008,soszynski2008optical, moskalik2009frequency}. In the latter study, a significant fraction ($\sim$9 \,per cent) of Cepheids with dominant radial first overtone and additional, likely non-radial mode (referred as FO-$\nu$ in their paper) was reported, based on analysis of the OGLE-II data. Two groups could be distinguished. In the first, the period of the additional variability is very close to the period of radial first overtone, $\Po$. In the second group, additional variability is always of higher frequency, with period ratios, $\Px/\Po\in (0.60,\, 0.65)$. In the following years more than 200 such stars were detected \citep[eg.][]{soszynski2010optical,soszynski2016multi,Suveges2018b}. In the Petersen diagram \citep[a plot of the shorter-to-longer period ratio vs. the longer period;][]{petersen1973masses} these stars form three distinct sequences. A detailed study of this group was conducted by \cite{smolec2016non}. They found that a noticeable fraction of these stars (35 \,per cent) show the signature of a power excess at the sub-harmonic frequency (1/2$\fx$). Interestingly, the majority of these are located in the middle sequence in the Petersen diagram. 

An analogous group of double-periodic pulsators with dominant radial first overtone and period ratios, $\Px/\Po$, in the $(0.60,\, 0.65)$ range exists also in first overtone RR Lyrae (RRc) stars \citep[eg.,][]{Gruberbauer2007MNRAS,Olech_Moskalik_2009A&A...494L..17O, moskalik2015kepler, netzel2019census}. The latest review of the physical interpretation of the non-radial modes in both 1O Cepheids and RRc stars via theoretical modeling is discussed by \cite{dziembowski2016}.  For Cepheids, the period ratios of $(0.60,\, 0.65)$ are proposed to be caused by harmonics of non-radial modes of moderate angular degrees ($\ell$) 7, 8 and 9. The two most populated sequences present in the Petersen diagram of RRc stars are explained through $\ell=8$ and 9 modes, while the third, middle sequence arises due to the linear combination of these two. 

\cite{Suveges2018b}, while investigating additional mode content in Magellanic Clouds OGLE data found yet another counterpart of a double-periodic group detected earlier in RRc stars by \cite{netzel2015discovery}. Dominant variability is due to radial first overtone and additional variability is of a longer period; characteristic period ratio is around 0.686 \citep[1.46-mode in][]{Suveges2018b}. The origin of this additional variability remains a mystery. One may conclude that there are parallels in empirical behavior between Cepheids and RR Lyrae stars, in terms of the presence of non-radial mode along with radial mode.

In contrast to RR Lyrae stars, which frequently exhibit large-amplitude quasi-periodic modulation of pulsation amplitude and phase, known as the Blazhko effect \citep{blazko1907mitteilung}, Cepheids still have a reputation of being regular pulsators. In the last decade, however, some of them have been found to exhibit periodic modulations in the radial modes as well. In addition to a long-known oddball, V437 Lyr \citep[eg.][]{Molnar2014MNRAS.442.3222M}, \cite{moskalik2009frequency} reported that a large fraction of double-mode, 1O+2O classical Cepheids have both radial modes periodically modulated, with large amplitude. Large-amplitude modulation was reported in a few single-mode 1O Cepheids \citep{soszynski2016multi}.  F-mode Cepheids appear more stable, but low-amplitude periodic modulation was reported in the only F-mode Cepheid in the {\it Kepler} field, V1154 Cyg \citep{Derekas2012MNRAS.425.1312D,Kanev2015EPJWC.10106036K,Derekas2017MNRAS.464.1553D}. Finally, \cite{smolec2017unstable} reported low-amplitude periodic modulation of pulsation in 51 F-mode Cepheids in the Magellanic Clouds, showing that for pulsation periods around 10 days the phenomenon may be quite common. Cycle-to-cycle variation of the radial velocities hinting at modulation were also reported \citep[eg.,][]{anderson2014tuning,Anderson2016MNRAS.463.1707A,Anderson2018pas6.conf..193A}.

Most of the above detections of additional periodicities and modulations were done for Magellanic Cloud Cepheids. The Galactic fields remain fairly unexplored. Information whether and how the above-discussed phenomena, their incidence rate and characteristics, depend on metallicity, or on a host population may be crucial for proposing and testing the models behind. This encouraged the present analysis -- search for additional periodicities in the OGLE Galactic Cepheids sample.

The structure of the paper is as follows. First, we present the data and analysis procedure in Section \ref{sec: Data and Analysis}. Results are presented in Section \ref{sec: Results}, followed by discussion in Section \ref{sec: Discussion}. In the end, we compile our conclusions from the study in Section \ref{sec: Summary and Conclusions}.

\section{DATA AND ANALYSIS}
\label{sec: Data and Analysis}
\begin{figure*}
\includegraphics[height=12.0cm,width=8cm]{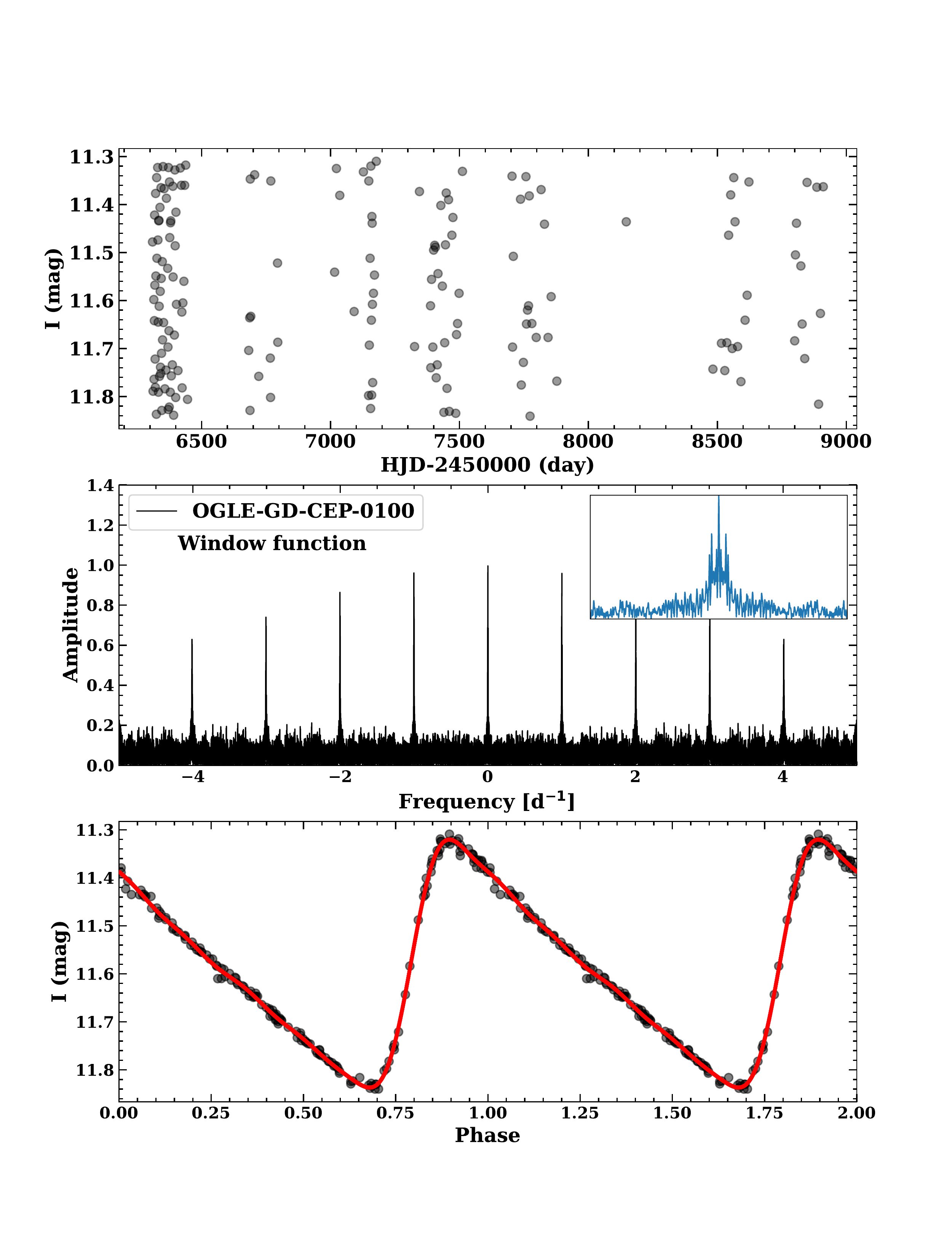}
\includegraphics[height=12.0cm,width=8cm]{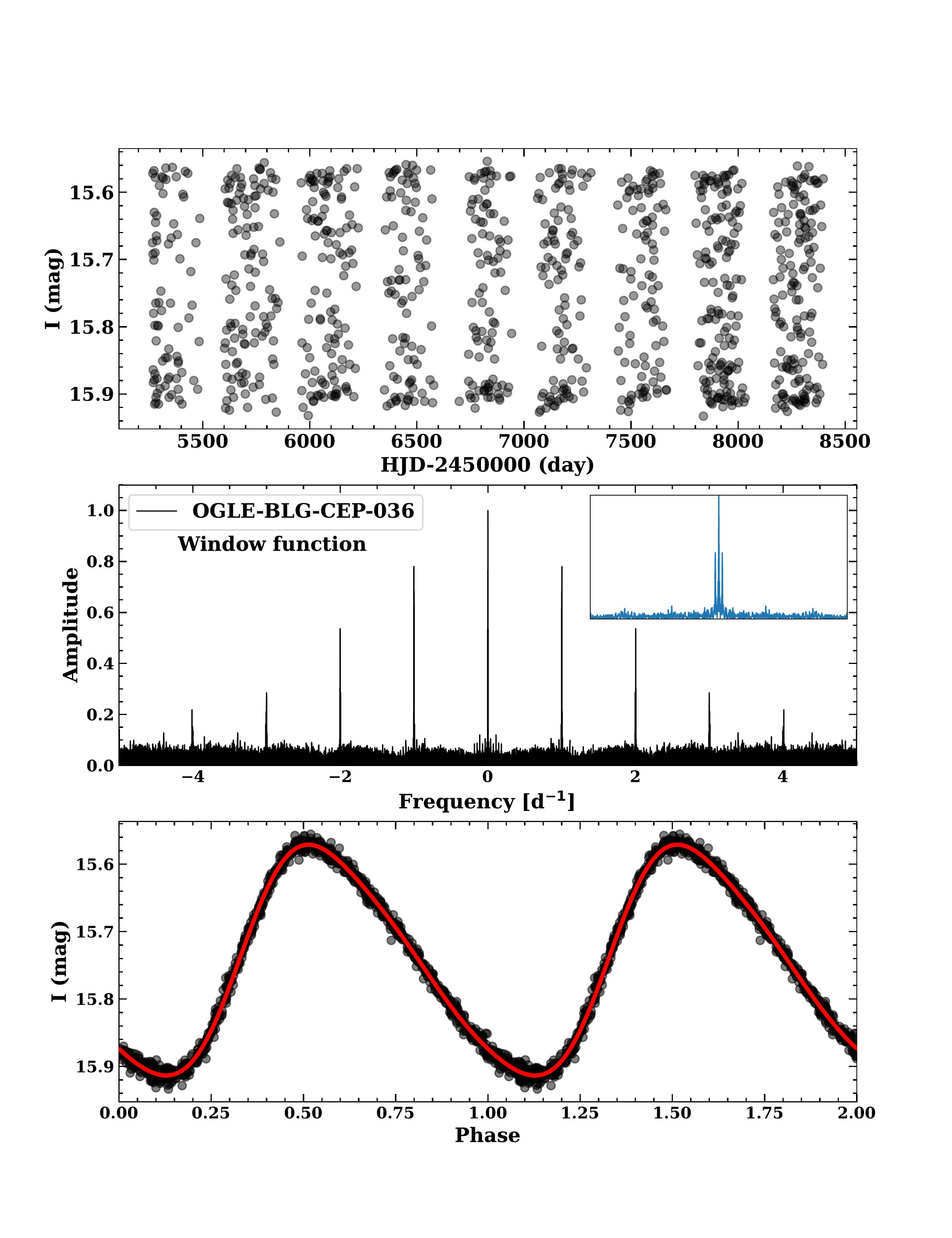}

\caption{The plot shows the representative time-series observations, window function and light curves to show the phase and temporal coverage for fundamental (Galactic disk) and first overtone (Galactic bulge) Cepheids. The subplot (in blue) in the middle panel shows the zoom-in at the window function centered at zero and the side peaks represent the yearly aliases.}
\label{fig: Representative light curves}
\end{figure*}

We conducted a frequency analysis study of Galactic disk and Galactic bulge Cepheids from Optical Gravitational Lensing Experiment (OGLE-IV) publicly available\footnote{\url{ftp://ftp.astrouw.edu.pl/ogle/ogle4}} photometric data \citep{udalski2015ogle,soszynski2017AcA....67..297S,Udalski2018AcA....68..315U,soszynski2020AcA....70..101S}. The analysis employs only I-band data as more sampling points are available, compared to the V-band. The fourth phase of the OGLE project, OGLE-IV, started in 2010 and is ongoing. OGLE-IV is ideal for investigation of mode content and of long-term pulsation behavior due to high-quality photometry (typical error of a single data point is a few mmag), long temporal baseline (up to 8 seasons) and high sampling rate (cadence of 19–60 minutes for inner Galactic bulge and 1–3 days for remaining bulge fields and Galactic disk).

We wrote a semi-automatic procedure to perform frequency analysis. As an initial step, we compute the Discrete Fourier Transform (DFT) of the time series to identify the principal/dominant frequency, $\nu_0$. Then the procedure fits a Fourier series as shown by the equation: 
\begin{equation}
\label{eq1}
m(t) = A_{0} + \sum_{k=1}^{N}A_{k} \sin(2\uppi k\nu_0t + \varphi_{k} ).
\end{equation}
Here, $A_{k}$ and $\varphi_{k}$ represent, respectively, amplitudes and phases of $k^{\rm{th}}$ order term. The order is chosen to satisfy $\frac{A_{k}}{\sigma(A_{k})} > 4 $ criteria, where $\sigma(A_{k})$ is the uncertainty in the amplitude.

During the automatic procedure, we remove 6-$\sigma$ outliers in the data (in the following manual analysis 4-$\sigma$ criterion is used) and also de-trend using low order polynomials to reduce possible season to season variation in the data, which may lead to undesired low frequency signals. Residuals are inspected for additional signals with S/N > 4 and based on this, stars are categorised into those potentially interesting and those in which the frequency spectra essentially contain noise.

After the automatic analysis is complete, the resulting inventory of interesting stars is analyzed manually. First, all resolved significant frequencies are included in the fit as independent frequencies and we look for any linear combination frequencies for subsequent inclusion in the sine series fit. Two signals are considered resolved, if their separation, $\Delta\nu$, fulfills the condition: $\Delta \nu>2/T$ where $T$ is the total time of the observation for a given star. Unresolved significant power at radial mode frequency may appear in the frequency spectrum due to the radial mode undergoing long-term phase/amplitude variations. Such significant unresolved power as a remnant in the frequency spectra increases the noise level and may hinder the detection of additional low-amplitude periodicities.
We get rid of the unresolved signals using the time-dependent pre-whitening technique \citep{moskalik2015kepler}, applied on a season-to-season basis.

Daily and yearly aliases in the frequency spectrum are inherent in the analysis of data from ground-based surveys such as OGLE. To illustrate this, data sampling, corresponding window function and light curve of two representative Cepheids are given by Fig.~\ref{fig: Representative light curves}. In the middle panels, zooms in the insets illustrate yearly aliases arising due to seasonal data sampling. At times (see for example window function for OGLE-GD-CEP-0100 in the left panel of Fig.~\ref{fig: Representative light curves}), the daily aliases are expected to be of comparable amplitude. Consequently, the identification of the {\it real} frequency of the periodic signal, based on peak amplitudes in the frequency spectra may be ambiguous. Such problematic cases that arise in our sample, will be discussed in detail in Section~\ref{subsec: Cepheids with double-mode radial pulsation}

Another type of unwanted signals that are sometimes detected in the frequency spectra are the ones that occur at close-to-integer frequencies and are most likely of instrumental origin. In OGLE data, a signal of the highest amplitude at $\approx$2$\,{\rm d}^{-1}$ is sometimes present in the frequency spectrum.

At times, there might be flux contamination of the primary source by the background stars. If the secondary is a variable star as well, it may introduce periodic signals in the frequency spectrum leading to erroneous results. In particular, we may suspect such contamination when we detect two signals with harmonics but detect no or only a few weak linear frequency combinations. To verify, we analyse flux instead of magnitude time series, as combination frequencies are not expected for two independent signals in flux data. This way we verified that  OGLE-BLG-CEP-067 and OGLE-BLG-CEP-0291 are not double-periodic stars, but contain two independent periodicities due to contamination. In OGLE-BLG-CEP-067, two independent periodicities detected in the star, 2.6107 d and 1.6924 d, are likely originating from two different Cepheids in the same line of sight.

{
\footnotesize
\begin{table}
\setlength{\tabcolsep}{-0.5pt}
\caption{Analysis summary of Cepheid sample in Galactic fields.}
\begin{tabular}{l|l|r|@{\hskip 0.2cm}r}

\hline
\hline
\textbf{Galactic Field}    & \textbf{Pulsation} & \textbf{ Raw Sample} & \textbf{For manual analysis}        \\
\hline
\hline
\textbf{DISK} & F & 1103 & 124  \\
 & 1O & 495 & 117   \\
 & F+1O & 45 & 45    \\
 & 1O+2O & 133  & 133  \\
 & 2O+3O & 1 & 1  \\
 & 1O+2O+3O & 5 & 5 \\
\hline
Total disk stars &  & 1782 & 425 \\         
\hline
\textbf{BULGE} & F & 94 & 19 \\
 & 1O & 56 & 32 \\
 & F+1O & 12 & 12 \\
 & 1O+2O & 19 & 19 \\
 & 2O+3O & 1 & 1 \\
 & 1O+2O+3O & 2 & 2 \\
\hline
Total bulge stars & & 184 & 85 \\
\hline
 
\end{tabular}
\label{tab: Analysis summary}
\end{table}
}

Since, the description of automatic procedure outlined above well suits the analysis of single-periodic stars and not multi-mode ones (F+1O; 1O+2O; 2O+3O; 1O+2O+3O), these stars were analyzed manually. Finally, starting from a raw sample of 1966 stars, our sample reduced to 510 Cepheids including both Galactic disk and bulge fields, for subsequent manual inspection. The complete analysis summary is given in Tab.~\ref{tab: Analysis summary}.

\section{RESULTS}
\label{sec: Results}

\begin{table*}
\caption{Properties of additional periodicities in stars classified in the OGLE collection as single-mode, F Cepheids. The columns below are as follows: OGLE ID (OGLE-GD/BLG-CEP-xxxx), radial fundamental mode period  (\Pf), period of the additional variability (\Px), ratio of radial fundamental mode and additional periodicity, $R_{\rm F}$, where $R_{\rm F}=\min(\Pf,\Px)/\max(\Pf,\Px)$, amplitude of the fundamental mode (\Af), amplitude of the additional variability (\Ax); signal to noise ratio of the additional variability (\sn$_{\rm x}$) and final column with remarks about the star (where pulsation classification is mentioned, `al' represents marking of lower alias signal, `ap' means additional periodicity, `det' means detrending of the time-series done, `nsF' means non-stationary fundamental mode, `comb' means combination frequencies detected). In some stars two possible solutions, sol 1/sol 2, of the frequency spectrum are given, arising due to alias ambiguity. In case one of the solutions seems more likely, it is marked with boldface font.}
\begin{tabular}{lllllllr}
\hline
\hline

\textbf{DISK F} & {} & {} & {} & {} & {} & {} & \\
OGLE ID & $\Pf$ (d) & $\Px$ (d) & $R_{\rm F}$ & $\Af$ (mag) & $\Ax$ (mag) & {$\sn_{\rm x}$} & Remarks\\
\hline
0425 & 2.604827(6) & 1.85392(8)  & {0.7117} & {0.1875} & {0.0071} & {4.2} & F+1O, al, comb  \\
0079 & 3.171777(5) &  4.3849(8)  & {0.7233} & {0.1548} & {0.0031} & {4.1} & det, ap \\

\hline
\hline
 \textbf{BULGE F} & {} & {} & {} & {} & {} & {} & \\
 OGLE ID & $\Pf$ (d) & $\Px$ (d) & $R_{\rm F}$ & $\Af$ (mag) & $\Ax$ (mag) & $\sn_{\rm x}$ & Remarks\\
 \hline
007  & 1.5455011(8) & 0.86369(4)  & 0.5588  & 0.2140 & 0.0030   & 3.7 & sol 1: F+2O, al\\

& 1.5455011(8) & 0.462842(9)  & 0.2995   & 0.2140 & 0.0037  & 4.5 & \textbf{sol 2: unknown} \\
                
040 & 3.022076(5) & 3.9104(4) & 0.7728  & 0.1364 & 0.0041 & 5.3  & det, ap, nsF  \\

\hline
\end{tabular}
\label{tab: F mode data table}
\end{table*}

\begin{table*}
\caption{Properties of additional periodicities in stars classified in the OGLE collection as double-mode, F+1O Cepheids. The columns below are as follows: OGLE ID (OGLE-GD/BLG-CEP-xxxx), radial fundamental mode period (\Pf), radial first overtone period (\Po), period of the additional variability (\Px), ratio of radial mode and additional periodicity, $R_{\rm F/1O}$, where $R_{\rm F/1O}=\min(\Pf/\Po,\Px)/\max(\Pf/\Po,\Px)$, amplitude of the radial fundamental mode (\Af), amplitude of the radial first overtone (\Ao), amplitude of the additional variability (\Ax), signal to noise ratio of the additional variability (\sn$_{\rm x}$) and final column with remarks about the star (where pulsation classification is mentioned, `ap' means additional periodicity.).}

\begin{tabular}{ll@{\hskip 0.2cm}l@{\hskip 0.2cm}l@{\hskip 0.2cm}l@{\hskip 0.2cm}l@{\hskip 0.2cm}l@{\hskip 0.2cm}l@{\hskip 0.2cm}l@{\hskip 0.2cm}l@{\hskip 0.1cm}r}
\hline
\hline
\textbf{DISK F+1O} &  &  &  &  &  &  &  &  \\
\textbf{OGLE ID} & {$\Pf$ (d)} & {$\Po$ (d)} & {$\Px$ (d)} & {$R_{\rm F}$} & {$R_{\rm 1O}$} & {$\Af$ (mag)} & {$\Ao$ (mag)} & {$\Ax$ (mag)} & {$\sn_{\rm x}$} & {Remarks} \\
\hline

0910 & 0.3614131(7) & 0.2731497(3) & 0.216995(1) & 0.6004  & 0.7944  & 0.0426   & 0.0650   & 0.0108  & 6.3  & F+1O+2O   \\
1743 & 9.8477(2) & 6.7516(1) & 8.251(2) & 0.8379  & 0.8182 & 0.0874 & 0.0709 & 0.0070  & 5.7 & ap\\

\hline
\hline
\end{tabular}
\label{tab: F+1O mode data table}
\end{table*}

\begin{table*}
\caption{Properties of additional periodicities in stars classified in the OGLE collection as double-mode, 1O+2O Cepheids. The columns below are as follows: OGLE ID (OGLE-GD/BLG-CEP-xxxx), radial first overtone mode period (\Po), radial second overtone period (\Poo), period of the additional variability (\Px), ratio of radial mode and additional periodicity, $R_{\rm 1O/2O}$, where $R_{\rm 1O/2O}=\min(\Po/\Poo,\Px)/\max(\Po/\Poo,\Px)$,
amplitudes of the first and second overtones ($\Ao$, $\Aoo$), amplitude of the additional variability (\Ax), signal to noise ratio of the additional periodicity (\sn$_{\rm x}$) and final column with remarks about the star (where pulsation classification is mentioned, `al' represents marking of lower alias signal, `ap' means additional periodicity, `det' means detrending of the time-series done, `tdp' means time dependent pre-whitening, `comb' means combination frequencies detected).}
\setlength{\tabcolsep}{-20pt}
\begin{tabular}{l@{\hskip 0.05cm}l@{\hskip 0.3cm}l@{\hskip 0.3cm}l@{\hskip 0.3cm}l@{\hskip 0.3cm}l@{\hskip 0.3cm}l@{\hskip 0.3cm}l@{\hskip 0.3cm}l@{\hskip 0.3cm}l@{\hskip 0.01cm}r}
\hline
\hline
\textbf{DISK 1O+2O} & {} & {} & {} & {} & {} & {} & {} & \\
 \textbf{OGLE ID} & {$\Po$ (d)} & {$\Poo$ (d)} & {$\Px$ (d)} & {$R_{\rm 1O}$} & {$R_{\rm 2O}$} & {$\Ao$ (mag)} & {$\Aoo$ (mag)} & {$\Ax$ (mag)} & {$\sn_{\rm x}$} & {Remarks} \\
\hline
0211 & 0.957678(4) & 0.77038(3) & 1.2938(2) & 0.7402 & 0.5954 & 0.1488 & 0.0138 & 0.0041 & 3.9 & F+1O+2O, al, tdp, nsO  \\
1638 & 0.945532(4) & 0.761901(9) & 1.30247(5)   & 0.7260 & 0.5850 & 0.0960 & 0.0302 & 0.0180 & 4.8 & F+1O+2O, det, al  \\
1711 & 0.4345698(7) & 0.349831(1) & 0.57440(1)  & 0.7566 & 0.6090 & 0.0921 & 0.0248 & 0.0179 & 4.0 & F+1O+2O, al  \\
1804 & 0.2410947(3) & 0.1917771(9) & 0.1596242(6) & 0.6621 & 0.8323 & 0.0806 & 0.0193 & 0.0181 & 6.4 & 1O+2O+3O, al \\ 
1730 & 0.262288(1) & 0.211827(1) & 0.35629(1)  & 0.7362 & 0.5945 & 0.0787 & 0.0596 & 0.0310 & 5.0 & F+1O+2O  \\
0106 & 0.2897632(2) & 0.231837(1) & 0.291936(3) & 0.9926 & 0.7941 & 0.1048 & 0.0103 & 0.0074 & 4.6 & ap \\
1610 & 0.327297(1) & 0.262079(4) & 0.159473(5) & 0.4872 & 0.6085 & 0.1224 & 0.0150 & 0.0061 & 4.1 & ap    \\

\hline
\hline

\textbf{BULGE 1O+2O} & {} & {} & {} & {} & {} & \\
\textbf{OGLE ID} & {$\Po$ (d)} & {$\Poo$ (d)} & {$\Px$ (d)} & {$R_{\rm 1O}$} & {$R_{\rm 2O}$} & {$\Ao$ (mag)} & {$\Aoo$ (mag)} & {$\Ax$ (mag)} & {$\sn_{\rm x}$} & {Remarks} \\
 
\hline
004 & 0.24004608(6) & 0.1902526(3) & 0.1520951(2) & 0.6336 & 0.7994 & {0.0538} & {0.0067} & 0.0054 & 12.9 & det, comb, ap  \\
                 &   &   & 0.1684677(5)  & 0.7018 & 0.8855 & & & 0.0029 & 7.2  & ap  \\
                 & &   & 0.1523386(5) & 0.6346 & 0.8007 & & & 0.0023 & 6.1  & ap \\
009 & 0.27256693(7) & 0.2167214(2) & 0.263823(2) & 0.9679 & 0.8215 & {0.0510} & {0.0136} & 0.0015 & 4.4  & det, comb \\
192 & 0.2494514(2) & 0.1967324(3) & 0.1222691(2) & 0.4902 & 0.6215 & {0.0346} & {0.0118} & 0.0062 & 7.3  & det, ap \\
                 &   &   & 0.2027133(7) & 0.8126 & 0.9705 &  &  & 0.0057 & 7.1  &  ap \\
\hline

\end{tabular}
\label{tab: 1O+2O mode data table}
\end{table*}

\begin{table*}
\caption{Properties of additional periodicities in stars classified in the OGLE collection as single-mode, 1O Cepheids. The columns below are as follows: OGLE ID (OGLE-GD/BLG-CEP-xxxx), radial first overtone mode period ($\Po$), period of the additional variability ($\Px$), ratio of radial first overtone period and additional variability, $R_{\rm 1O}$, where $R_{\rm 1O}=\min(\Po,\Px)/\max(\Po,\Px)$, amplitude of the first overtone ($\Ao$), amplitude of the additional variability ($\Ax$), amplitude ratio ($\Ax/\Ao$), signal to noise ratio of the additional variability ($\sn_{\rm x}$) and final column with remarks about the star (where pulsation classification is mentioned, `al' represents marking of lower alias signal, `det' means detrending of the time-series done, `nsO' means non-stationary first overtone mode, `ns' refers to non-stationary additional signal/power excesses, `tdp' means time dependent pre-whitening, `comb' means combination frequencies detected).  In some stars two possible solutions, sol 1/sol 2, of the frequency spectrum are given, arising due to alias ambiguity. In case one of the solutions seems more likely, it is marked with boldface font.}
\begin{tabular}{llllllllr}
\hline
\hline
\textbf{DISK 1O} & {} & {} & {} & {} & {} & \\
OGLE ID & $\Po$ (d) & $\Px$ (d) & $R_{\rm 1O}$ & $\Ao$ (mag) & $\Ax$ (mag)& $\Ax/\Ao$  & $\sn_{\rm x}$ & Remarks\\
\hline
0032 & 0.3905313(4) & 0.312673(4) & 0.8006 & 0.1167 & 0.0076 & 0.0649 & 5.2 & \textbf{sol 1: 1O+2O} \\
 & 0.3905314(4) & 0.238040(2) & 0.6095 & 0.1169 & 0.0078 & 0.0663 & 5.2 & sol 2: F+2O, al \\
0847 & 0.547452(7) & 0.44155(3) & 0.8066 & 0.1077 & 0.0265 & 0.2465 & 4.7 & 1O+2O \\ 
 0936 & 0.3095034(6) & 0.247183(3) & 0.7986 & 0.0732 & 0.0101 & 0.1377 & 4.2 & 1O+2O, al \\
 1220 & 0.507044(2) & 0.68168(5) & 0.7438 & 0.0735 & 0.0089 & 0.1210 & 4.3 & sol 1: F+1O \\
 & 0.507044(1) & 0.404911(4) & 0.7986 & 0.0735 & 0.0090 & 0.1221 & 4.3 & sol 2: 1O+2O, al \\
 1479 & 1.7060(1) & 1.3698(2) & 0.8029 & 0.1022 & 0.0218 & 0.2133 & 4.7 & 1O+2O, al \\
1620 & 0.2345349(3) & 0.187493(2) & 0.7994 & 0.1080 & 0.0087 & 0.0807 & 6.4 & 1O+2O, al, comb \\
1753 &  0.2540406(3) & 0.202664(2) & 0.7978 & 0.0841 & 0.0100 & 0.1187 & 5.1 & \textbf{sol 1: 1O+2O, al} \\
  & 0.2540404(4) & 0.168436(3) & 0.6630 & 0.0842 & 0.0114 & 0.1351 & 5.2 & sol 2: 1O+3O \\
1806 & 0.2551434(9) & 0.203460(3) & 0.7974 & 0.0713 & 0.0134 & 0.1885 & 4.2 &\textbf{ sol 1: 1O+2O, al} \\
 & 0.2551442(5) & 0.168967(2) & 0.6623 & 0.0705 & 0.0127 & 0.1795 & 4.2 & sol 2: 1O+3O \\
0076 & 3.01033(2) & 2.1462(2)  & 0.7130 & 0.1081 & 0.0051 & 0.0473 & 5.1 & sol 1: F+1O \\
 &  3.01033(2) & 1.8629(2) & 0.6188 & 0.1080 & 0.0051 & 0.0469 & 4.9 & \textbf{sol 2: Group 1}, al \\
 0420 & 1.194529(2) & 0.71745(2) & 0.6006 & 0.1520 & 0.0046 & 0.0302 & 4.0 & \textbf{sol 1: Group 1}, al \\
 & 1.194528(2) & 1.60766(1) & 0.7430 & 0.1523 & 0.0046 & 0.0305 & 3.9 & sol 2:  F+1O, al \\
0126 & 3.27317(2) & 1.9765(1) & 0.6038 & 0.0922 & 0.0060 & 0.0649 & 4.2 & Group 1, nsO \\
0041 & 2.49740(2) & 1.4651(1) & 0.5866 & 0.0928 & 0.0045 & 0.0488 & 3.7 & Group 1, al \\ 
0632 & 3.66766(5) & 4.689(2) & 0.7821 & 0.0900 & 0.0046 & 0.0515 & 4.7 & Group 2, al, ns \\ 
0793 & 4.28649(3) & 2.6763(4) & 0.6244 & 0.1160 & 0.0036 & 0.0313 & 3.5 & Group 1, det, tdp, nsO \\
0932 & 4.25461(3) & 2.7066(3) & 0.6362 & 0.1300 & 0.0049 & 0.0374 & 4.5 & Group 1, al, det \\
1346 & 1.74826(3) & 1.0952(3) & 0.6264 & 0.1078 & 0.0061 & 0.0567 & 4.0 & Group 1, al \\
1308 & 3.6996(6) & 2.1815(8) & 0.5897 & 0.1245 & 0.0322 & 0.2588 & 5.6 & Group 1 \\
1314 & 2.41598(5) & 1.5012(3) & 0.6214 & 0.1158 & 0.0073 & 0.0629 & 6.3 & Group 1 \\
1668 & 2.68862(1) & 1.6817(1) & 0.6255 & {0.1114} & 0.0045 & 0.0403 & 4.0 & Group 1 \\
&  & 3.1564(3) & 0.8518 &  & 0.0069 & 0.0616 & 4.5 & Group 2, ns \\
0850  & 3.67494(3) & 5.4984(10) & 0.6684 & 0.0879 & 0.0046 & 0.0527 & 4.3 & ap \\
  &  &  2.2785(3) & 0.6200 & & 0.0029 & 0.0335 & 3.5 & Group 1 \\
0475 & 4.31162(3) & 5.7684(9) & 0.7475 & 0.1393 & 0.0064 & 0.0462 & 4.3 & Group 2, comb, ns \\
0546 & 5.17904(4) & 6.629(2) & 0.7812 & 0.0867 & 0.0046 & 0.0530 & 4.6 & Group 2, det, ns \\
1332 & 1.86971(5) & 1.1658(2) & 0.6235 & 0.1073 & 0.0047 & 0.0440 & 4.1 & Group 1 \\

 &  & 2.442(2) & 0.7655 &  & 0.0050 & 0.0464 & 4.4 & Group 2, ns \\
1280 & 3.07538(1) & 3.951(2)& 0.7783 & 0.0944 & 0.0049 & 0.0517 & 4.1 & Group 2 \\
0863 & 6.47964(5) & 7.812(3) & 0.8294 & 0.1149 & 0.0039 & 0.0338 & 4.6 & Group 2, det \\
1193 & 5.10932(8) & 6.894(2) & 0.7412 & 0.0772 & 0.0055 & 0.0718 & 5.1 & Group 2, al, det \\
1339 & 3.30572(9) & 4.176(4) & 0.7917 & 0.0919 & 0.0046 & 0.0499 & 4.0 & Group 2, al, ns \\

\hline
\hline
\textbf{BULGE 1O} & {} & {} & {} & {} & {} & \\
 OGLE ID & $\Po$ (d) & $\Px$ (d) & $R_{\rm 1O}$ & $\Ao$ (mag) & $\Ax$ (mag) & $\Ax/\Ao$  & $\sn_{\rm x}$ &Remarks\\
\hline
024 & 0.35542535(4) & 0.199454(1) & 0.5612 & 0.1656 & 0.0016 & 0.0097 & 4.3 & det, tdp, ap \\
027 & 0.29652811(7) & 0.276636(2) & 0.9329 & 0.0541 & 0.0010 & 0.0187 & 5.2 & nsO, tdp \\
 &  & 0.1047693(3) & 0.3533 & {} & 0.0011 & 0.0206 & 6.1 & ap \\
056 & 4.76894(2) & 6.241(1) & 0.7642 & 0.0936 & 0.0023 & 0.0250 & 4.2 & Group 2,det, tdp, ns\\
196 & 0.2535573(2) & 0.2558523(4)  & 0.9878 & 0.0747 & 0.0175 & 0.2343 & 13.2 & ap \\

\hline
\end{tabular}
\label{tab: 1O mode data table}
\end{table*}

In our sample we have detected new candidates for double-mode and triple-mode radial pulsation (Section~\ref{subsec: Cepheids with double-mode radial pulsation}) and several candidates for double-periodic pulsators in which the detected additional periodicity cannot be associated with radial mode (Section~\ref{subsec: Cepheids with double periodic pulsation} and \ref{subsec: Additional periodicities in other radial mode sample}). We have also discovered periodic modulation of pulsation in a few classical Cepheids including F-mode pulsator, 1O pulsator and double-mode radial pulsator, F+1O (Section~\ref{subsec: Periodic modulation of pulsation}). Detection of modulation in F+1O double-mode Cepheid is the first reported in literature whereas the modulation in F-mode is a Galactic detection, adding to the scarce collection of such candidates in the literature. 

\subsection{New detection of radial multi-mode Cepheids}
\label{subsec: Cepheids with double-mode radial pulsation}

In several Cepheids, we have detected additional periodicity that can be attributed to radial mode. Two stars were originally classified as single-mode F-mode pulsator, ten stars as single-mode 1O mode pulsators (candidates for new double-mode Cepheids), one star as double-mode F+1O pulsator, and five stars as double-mode 1O+2O pulsators (candidates for new triple-mode Cepheids). Data for these stars are collected in the top sections of Tabs~\ref{tab: F mode data table}--\ref{tab: 1O mode data table} for stars originally classified as F, F+1O 1O+2O and 1O pulsators, respectively. The columns in the tables contain star id (OGLE-GD/BLG-CEP-xxx), periods of the already classified radial modes ($P_{\rm F/1O/2O}$) and of the additional detection ($\Px$), corresponding period ratios -- shorter to longer period, amplitude of radial modes ($A_{\rm F/1O/2O}$), amplitude of additional detection ($\Ax$), signal-to-noise ratio for the new detection ($\sn_{\rm x}$) and remarks.

In 10 stars of the OGLE single-mode 1O sample we have detected additional periodicities that may be classified as radial modes, ranging from fundamental to radial third overtone. For some of these stars alternate interpretation is possible which is mentioned as an added row against the star in Tab.~\ref{tab: 1O mode data table}. Such multiple interpretation are due to ambiguities in characterizing {\it correct} peak in the frequency spectrum due to daily aliases of comparable height.

\begin{figure*}
\centering
\includegraphics[height=10cm,width=16cm]{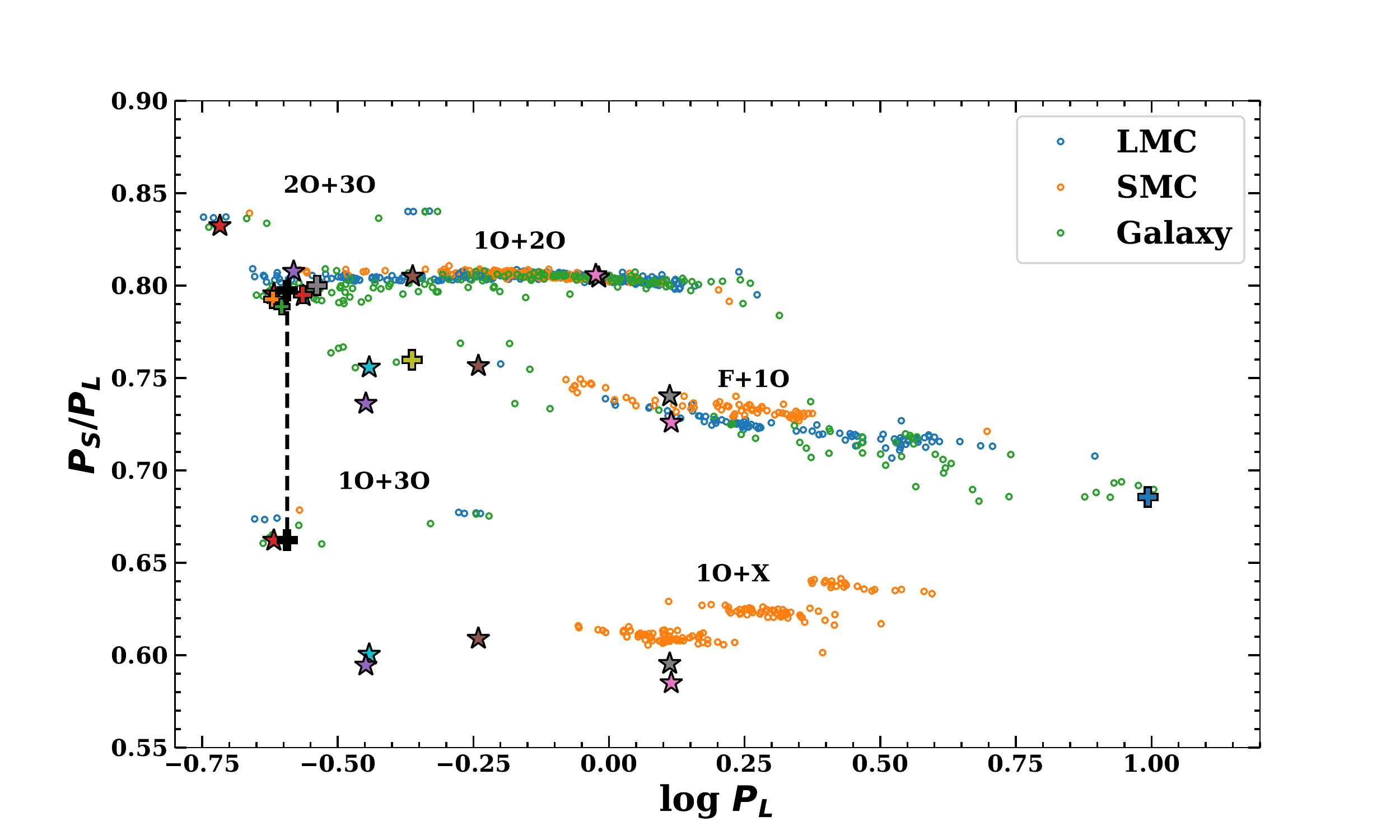}
\caption{Petersen diagram showing new detections of double-mode (plus symbol) and triple mode (star symbol) candidates (same colour denotes different modes for the same star). Hollow circles are OGLE-IV Cepheids from LMC (blue) SMC (orange) and Galaxy (green). Black dotted line connects the two possible double-mode solutions for OGLE-GD-CEP-1806 as mentioned in Sect.~\ref{subsec: Cepheids with double-mode radial pulsation}.}
\label{fig: Petersen_1O2O}
\end{figure*}

\begin{figure}
\centering
\includegraphics[height=9cm,width=9cm]{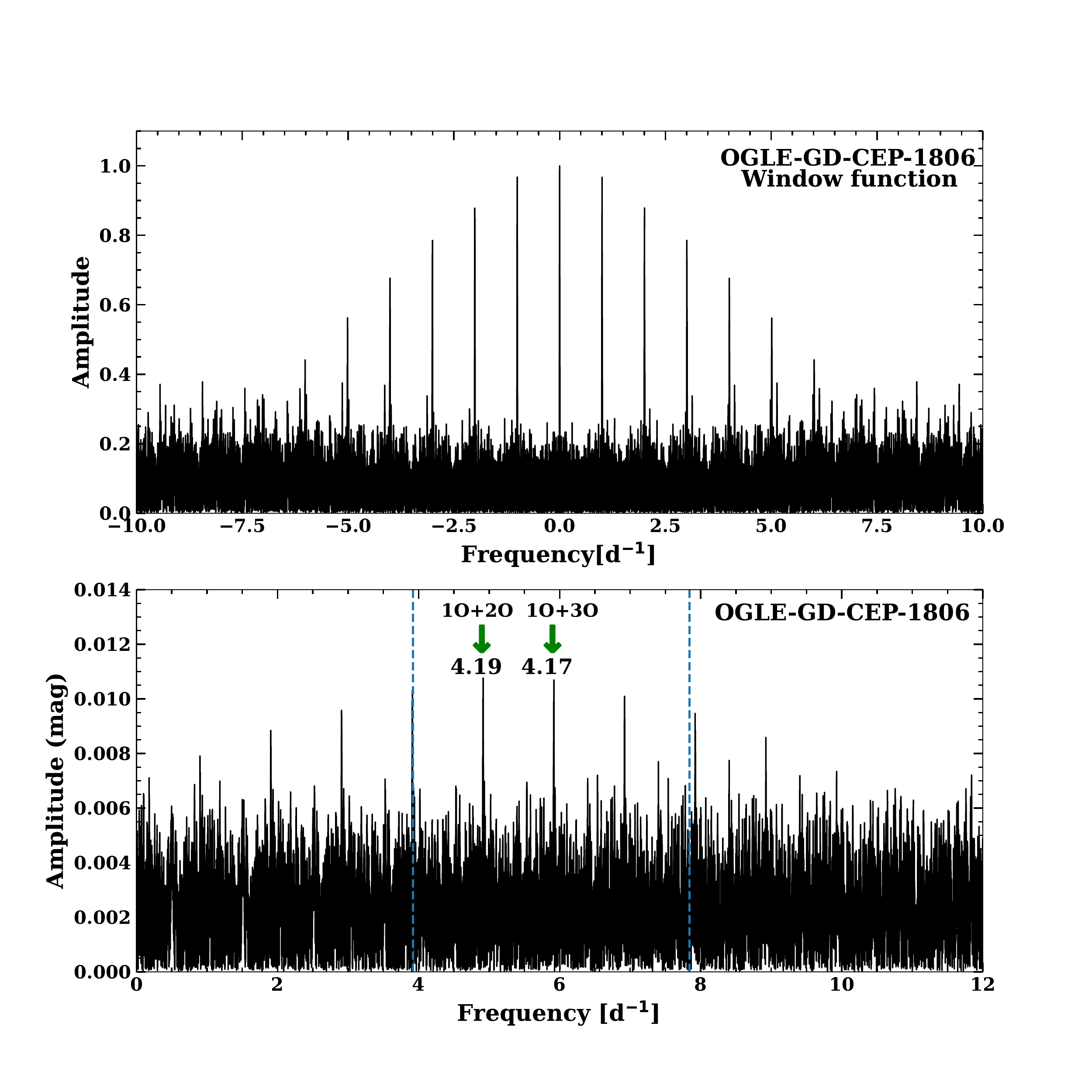}
\caption{Frequency spectra for OGLE-GD-CEP-1806. Upper panel shows a window function. Lower panel shows frequency spectrum pre-whitened with radial first overtone and its harmonic (blue dashed lines). Two daily aliases are marked with their respective $\sn$ and interpretation.}
\label{fig: OGLE-GD-CEP-1806}
\end{figure}

One of the representative of the ambiguities that arise in radial mode classification is well captured for OGLE-GD-CEP-1806 ($\Po\approx 0.255$\,d). After pre-whitening the first overtone frequency and its harmonic -- see Fig.~\ref{fig: OGLE-GD-CEP-1806} -- the highest peak ($\sn=4.19$ may be interpreted as due to radial 2O, with $\Px/\Po=0.7974$, whereas a bit lower daily alias ($\sn=4.17$) may be interpreted as due to radial 3O ($\Px/\Po= 0.6623$). Both solutions are illustrated in the Petersen diagram in Fig.~\ref{fig: Petersen_1O2O} (crosses connected with the dashed line) and fall well withing the 1O+2O or 1O+3O progressions. As window function in the top panel of Fig.~\ref{fig: OGLE-GD-CEP-1806} indicates, the consecutive daily aliases are expected to be of similar height. Even though both solutions are likely, our preference is inclined to double-mode 1O+2O pulsation, as this pulsation type involving consecutive radial modes is much more frequently detected as compared to the 1O+3O pulsation. A similar example was seen in OGLE-GD-CEP-1753 ($\Po\approx 0.254$\,d) from 1O sample, with two radial double-mode classification possibilities. The first one being 1O+2O ($\sn=5.11$, $\Px/\Po=0.7978$; preferred solution) pulsation and the other possibility as 1O+3O ($\sn=5.17$, $\Px/\Po=0.6630$). 

OGLE-GD-CEP-0032 (classified as 1O, with $\Po\approx0.391$\,d) presents yet another case where two solutions are possible. The first case indicates the presence of second overtone radial mode ($\Px/\Po=0.8006$, $\sn=5.2$). By selecting the other alias of comparable height, $\sn=5.2$, we get $\Px/\Po=0.6095$, which does not point towards double-mode radial pulsation, unless the primary frequency corresponds to radial fundamental mode instead of radial 1O. Then, the period ratio points towards F+2O double-mode radial pulsation. The primary periodicity is of too low amplitude to firmly identify it as due to radial F or radial 1O based on light curve shape. Our preference goes to 1O+2O solution that agrees with primary OGLE classification and represents a more common class of double-mode pulsators.

OGLE-GD-CEP-1220  (classified as 1O, $\Po\approx0.507$\,d) is another star for which two radial double-mode scenarios are possible based on which peak is selected after pre-whitening with the first overtone frequency. If higher frequency alias is selected ($\sn=4.3$), the period ratio, $\Px/\Po=0.7986$, points toward 1O+2O pulsation. If lower frequency alias is selected ($\sn=4.3$) the period ratio of $\Po/\Px=0.7438$ points toward F+1O double-mode pulsation.

Fundamental mode sample analysis gave a couple of detection of double-mode candidates. In OGLE-GD-CEP-0425 ($\Pf\approx2.605$\,d), we detect an additional periodicity of $1.854$\,d, which can be associated with radial first overtone as indicated by period ratio of $\Px/\Pf= 0.7117$ (see Tab.~\ref{tab: F mode data table}). Another originally classified F-mode candidate OGLE-BLG-CEP-007 ($\Pf\approx1.546$\,d), after pre-whitening with all possible F-mode harmonics, gave two possible solutions. The first is when the lower alias ($\sn=3.7$) of period 1.854 d is considered; it may correspond to second overtone radial mode which makes the star a candidate F+2O double-mode Cepheids. For the second solution ($\sn=4.5$), the period is shorter than in the above solution ($\Px\approx0.463$\,d) and it does not corresponds to any radial mode interpretation.

In the OGLE F+1O sample we have identified one star, OGLE-GD-CEP-0910, that may, in fact, be triple-mode radial pulsator. The candidate and its pulsation period/amplitude data is provided in Tab.~\ref{tab: F+1O mode data table}. In this Cepheid, after pre-whitening with fundamental and first overtone radial modes we identify a significant ($\sn=6.3$) detection of a third periodicity ($\Px=0.217$\,d). The period ratios ($\Px/\Pf=0.6004$ and $\Px/\Po=0.7944$) indicate that it is likely a F+1O+2O candidate. Moreover, on the Petersen diagram, this star lies at the boundary of multi-mode delta Scuti and classical Cepheids. Previously cataloged Cepheids of triple-mode nature in OGLE database are rare with only 2 detections in Galactic disk and none in the bulge \citep{soszynski2017AcA....67..297S}.

In 1O+2O inventory, we found 5 triple-mode candidates. The periods and amplitudes for the presumably additional radial modes in the sample are provided in Tab.~\ref{tab: 1O+2O mode data table}. All of these triple-mode detections were from the Galactic disk sample. OGLE-GD-CEP-1804 presents one such case likely with third radial overtone mode present as period ratios $\Px/\Po=0.6621$ and $\Px/\Poo=0.8323$ indicate. Moreover, in three Cepheids, OGLE-GD-CEP-0211, OGLE-GD-CEP-1638 and OGLE-GD-CEP-1711, the additional periodicity is of longer period; corresponding period ratios are $\Po/\Px=(0.7402,\, 0.7260,\, 0.7566)$ and $\Poo/\Px=(0.5954,\, 0.5850,\, 0.6090)$. Hence, our analysis points towards them being triple-mode candidates (F+1O+2O) with fundamental radial mode periods of $1.294$, $1.302$ and $0.574$\,d, respectively. There is another candidate star, OGLE-GD-CEP-1730, for F+1O+2O class of pulsation with $\Po/\Px=0.7362$ and $\Poo/\Px=0.5945$. However, in this particular candidate, the presence of a third radial mode is only one of the possible solutions depending on the alias selected. In the second solution, we have an unresolved peak at radial first overtone frequency, which may arise due to non-stationary nature of the radial 1O.

For last radial double-mode group, 2O+3O, and triple mode-mode group, 1O+2O+3O, in raw sample we had 2 and 7 Cepheids respectively. These were analyzed manually and eventually no traces of additional periodicities, that could be associated with radial modes, were found in these samples.

\subsection{Double-periodic pulsation with $\Px/\Po\in(0.60,\, 0.65)$ and associated variability}
\label{subsec: Cepheids with double periodic pulsation}

\begin{figure*}
\centering
\includegraphics[height=10cm,width=16cm]{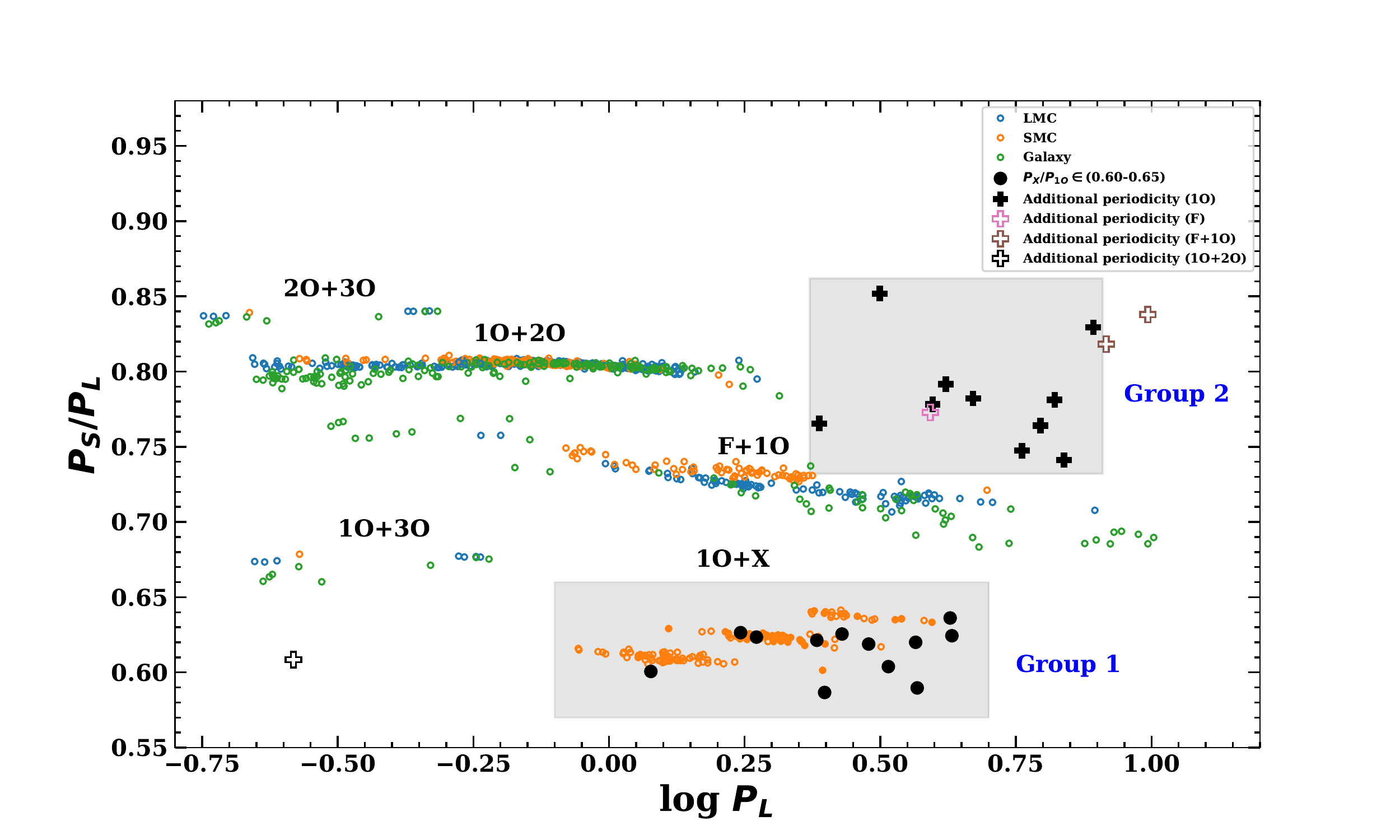}
\caption{Petersen diagram showing candidates with period ratios in the $(0.60,\, 0.65)$ range (black filled circle; marked as `Group 1') and additional periodicity detection (black plus symbol) in the first overtone sample. Hollow plus symbol denote our detection of additional periodicity in radial F (pink); F+1O (brown) and 1O+2O (black) pulsators. Literature detection are taken from SMC \citep{smolec2016non} (orange circle) for 1O+additional non-radial periodicity Cepheids and other double-mode Cepheids from OGLE-IV database. Colour scheme as per metallicity environment is as follows: LMC (blue hollow circle), SMC (orange hollow circle) and Galaxy (green hollow circle).}
\label{fig: 1O+X and subharmonic}
\end{figure*}

\begin{figure}
\centering
\includegraphics[height=6cm,width=9cm]{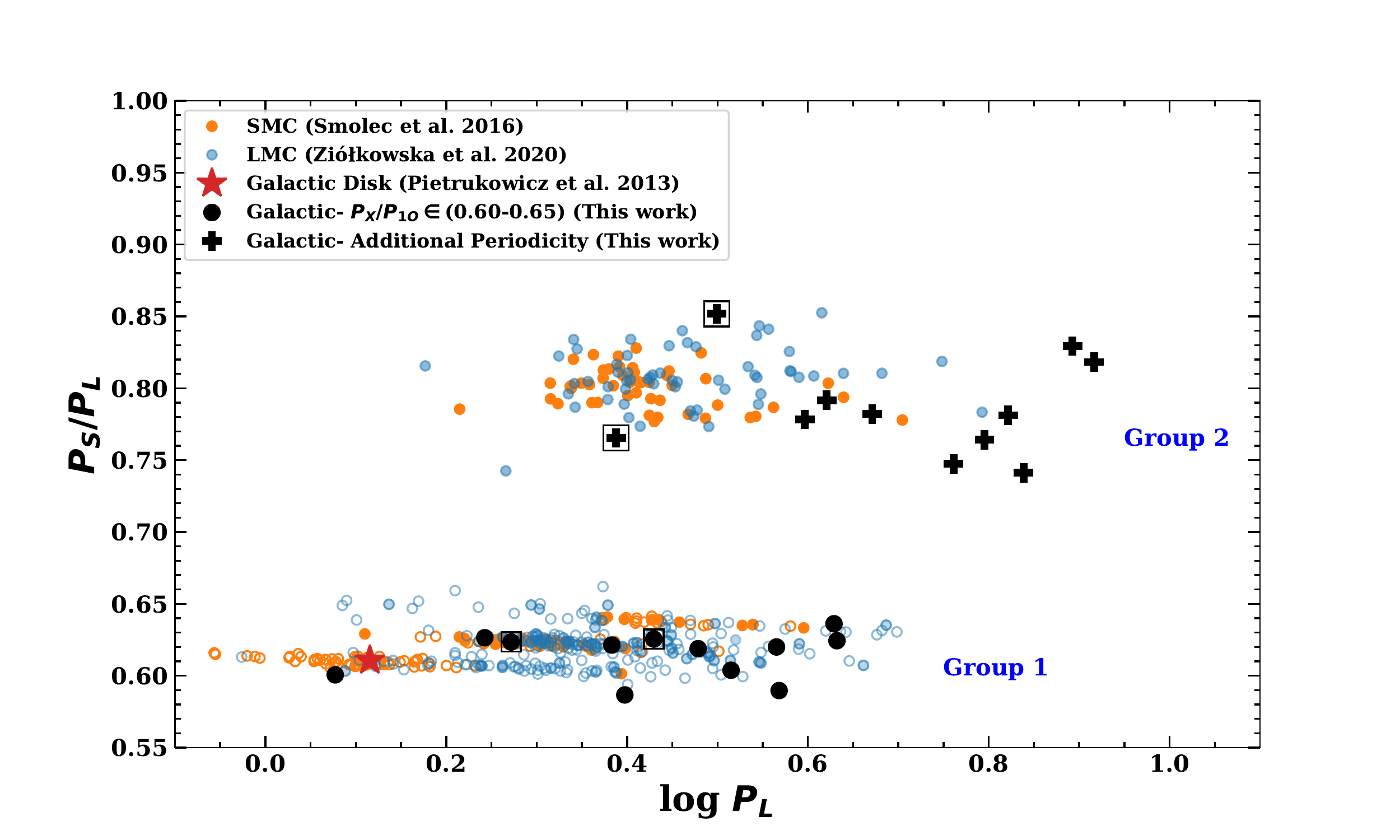}
\caption{Petersen diagram showing Galactic sample candidates with period ratios in the $0.60-0.65$ range (`Group 1'; black filled circles) and possibly, detection of sub-harmonic signal (`Group 2', black plus symbols). The symbol enclosed in black square represents OGLE-GD-CEP-1668 and OGLE-GD-CEP-1332 with simultaneous presence of periodicity in both groups. The literature detections are taken from SMC \citep[][orange circles]{smolec2016non}, LMC \citep[][blue circles]{Ziolkowska2020past.conf...75Z} and \citet{Pietrukowicz2013AcA....63..379P}.}

\label{fig: Zoom_non_radial}
\end{figure}

\begin{figure*}
\centering
\includegraphics[height=5cm,width=8cm]{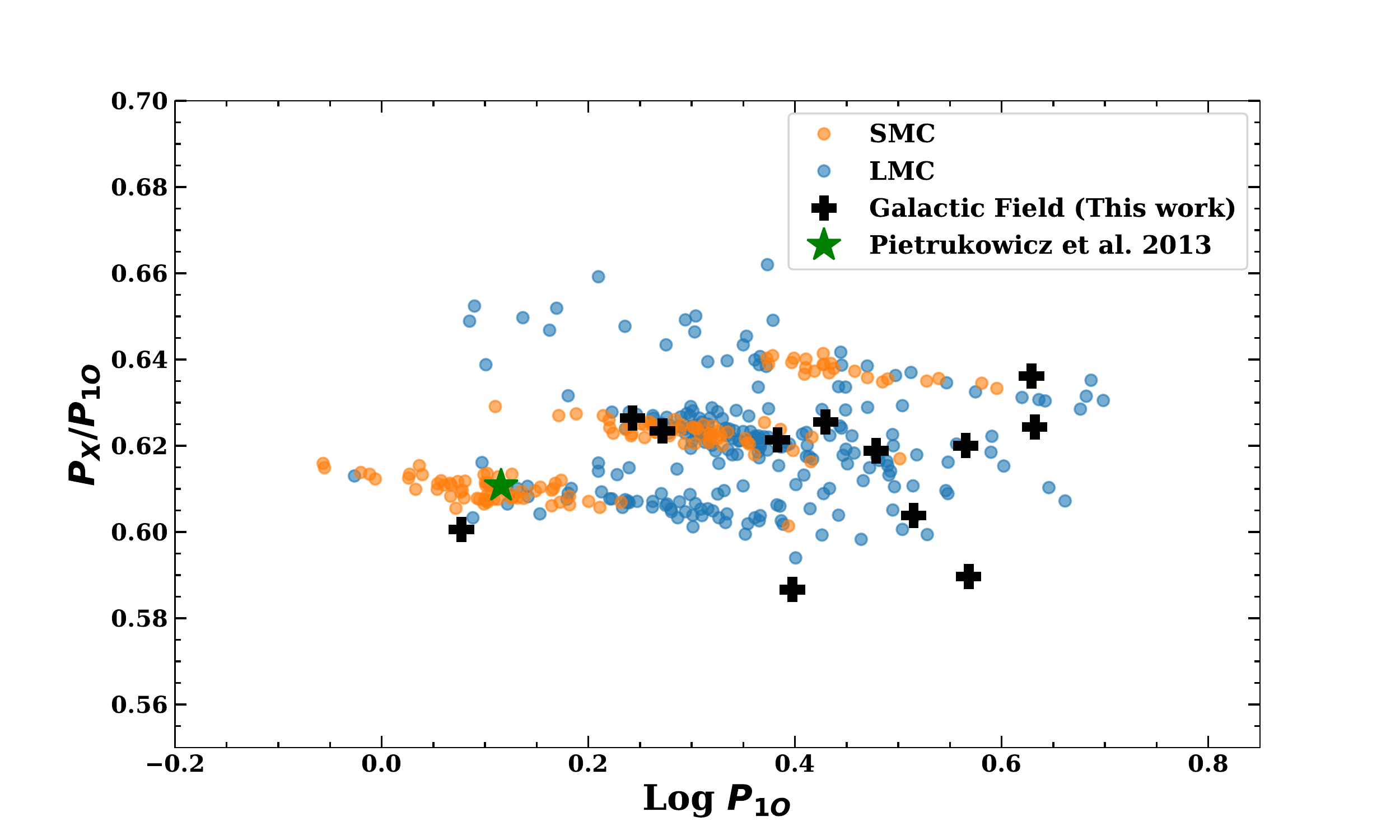}
\hspace{-0.7cm}
\includegraphics[height=5cm,width=8cm]{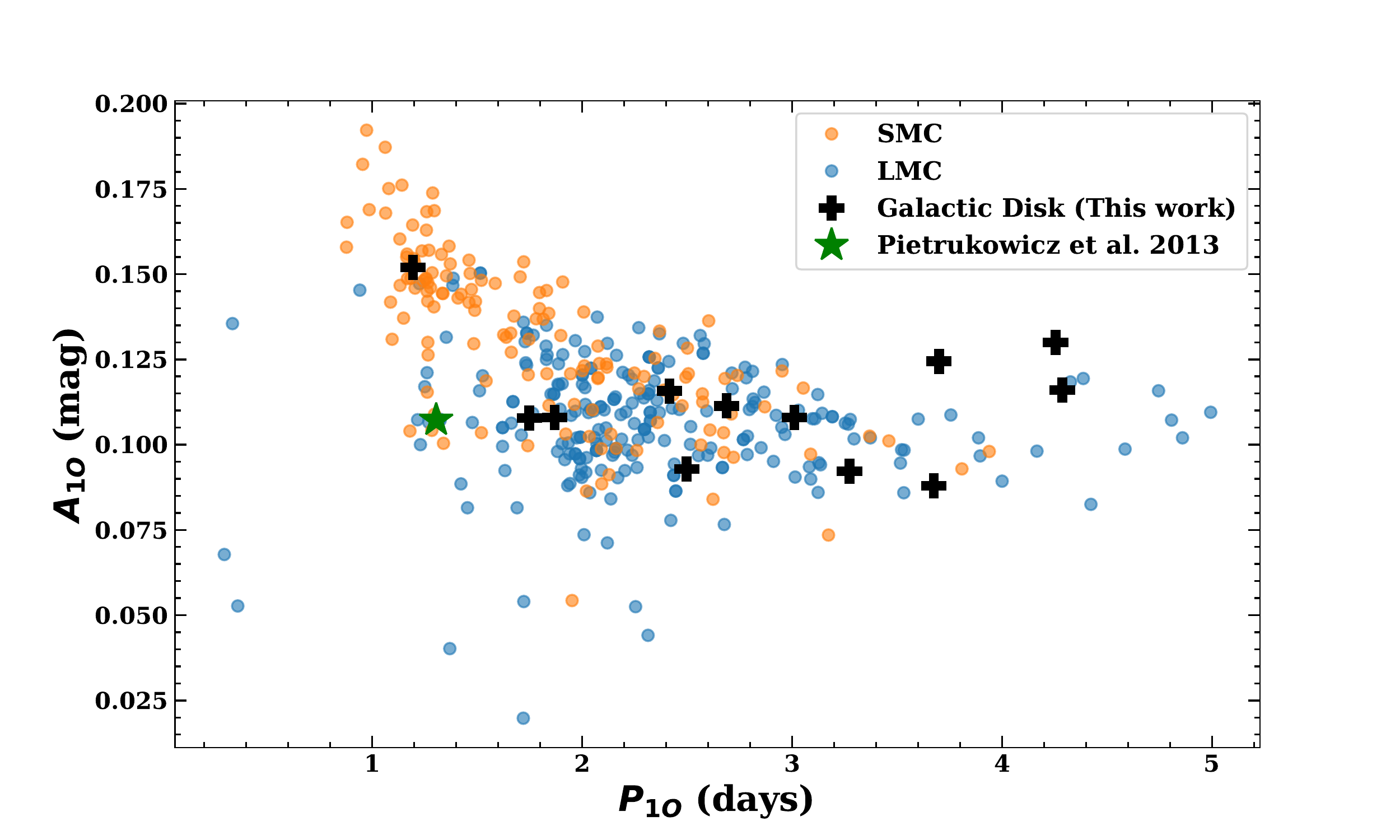}
\includegraphics[height=5cm,width=8cm]{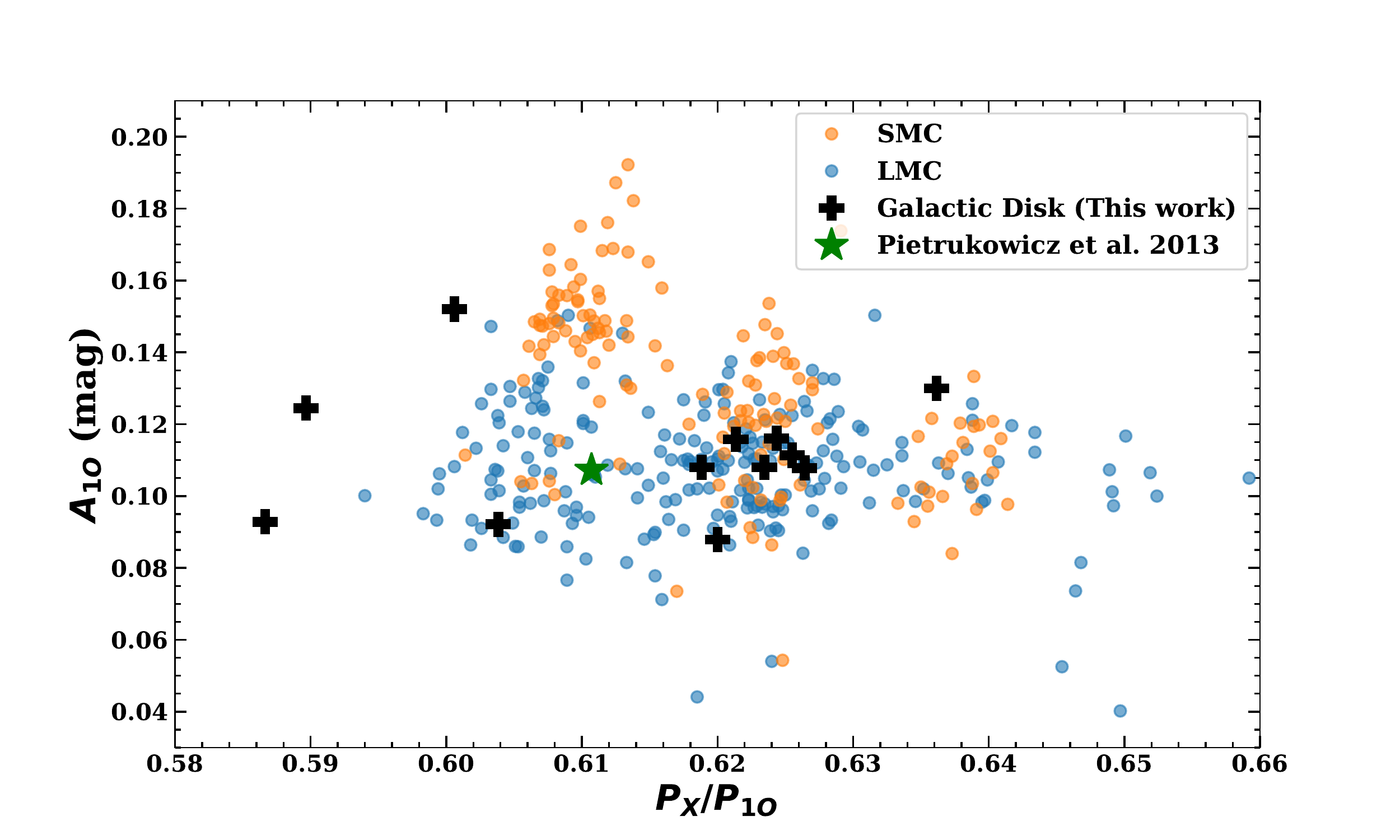}
\hspace{-0.7cm}
\includegraphics[height=5cm,width=8cm]{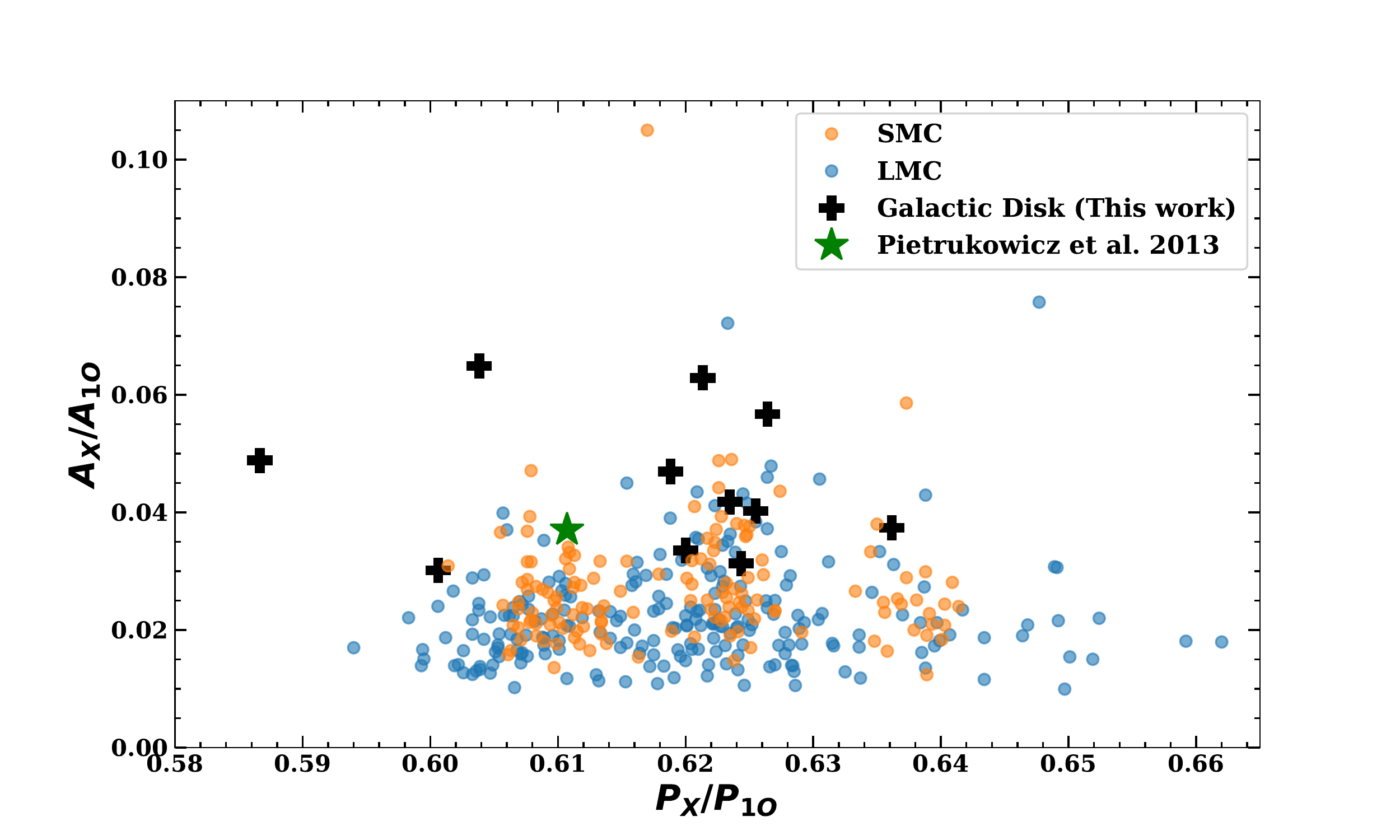}
\caption{Comparison of periods and amplitudes of non-radial first overtone Cepheids (black plus symbols) with similar classes of stars from SMC \citep[orange circles,][]{smolec2016non}, LMC \citep[blue circles,][]{Ziolkowska2020past.conf...75Z} and Milky-Way \citep[green star symbol,][]{Pietrukowicz2013AcA....63..379P}}
\label{fig: non-radial first overtone}
\end{figure*}

In this section we focus on 12 stars in which additional periodicity corresponds to $\Px/\Po\in (0.60,\, 0.65)$ (`Group 1' in Figs~\ref{fig: 1O+X and subharmonic} and \ref{fig: Zoom_non_radial}) and on stars in which additional periodicity seems to be directly linked to the same group (`Group 2' in Figs~\ref{fig: 1O+X and subharmonic} and \ref{fig: Zoom_non_radial}). Cepheids with additional periodicities in the $\Px/\Po\in (0.60,\, 0.65)$ range were detected earlier in the LMC \citep{moskalik2009frequency,soszynski2008optical}
and the SMC \citep{soszynski2010optical}; in the latter system, stars form three clear and distinct sequences in the Petersen diagram (orange open circles in Fig.~\ref{fig: 1O+X and subharmonic}).  Recently, \cite{Ziolkowska2020past.conf...75Z} analyzed 85\,per cent of the LMC 1O sample in OGLE-IV database and reported 277 stars with additional low-amplitude periodicities, a fraction of which corresponds to period ratio mentioned above. Later, \cite{Pietrukowicz2013AcA....63..379P} detected additional periodicity in one Galactic Cepheid (OGLE-GD-CEP-0001), belonging to the bottom sequence, from OGLE-III data. These double-periodic stars from three metallicity fields are plotted in the Petersen diagram in Fig.~\ref{fig: Zoom_non_radial}. The basic pulsation properties for stars from our Galactic sample are presented in Tab.~\ref{tab: 1O mode data table}. A more detailed comparison with the SMC and LMC samples, as analysed by \cite{smolec2016non} and \cite{Ziolkowska2020past.conf...75Z}, respectively, is illustrated in Fig.~\ref{fig: non-radial first overtone} in which the panels show the Petersen diagram, a plot of $\Ao$ vs. $\Po$ and plots of $\Ao$ and amplitude ratio, $\Ax/\Ao$ vs. the period ratio, $\Px/\Po$. Based on these plots we conclude that only one of our detections fits well within the top sequence, whereas likely four correspond to the bottom sequence, provided the sequence is extended to significantly longer first overtone periods. Comparatively, more stars (7) are found to fit the middle sequence. There is a larger spread in period ratios in our sample as one goes to longer first overtone periods. The amplitude ratio ($\Ax/\Ao$) for Galactic Cepheids is higher than for both LMC and SMC stars in all three sequences.

The analyses of \cite{smolec2016non} and \cite{Ziolkowska2020past.conf...75Z} \citep[see also][]{Suveges2018b} show that in a significant fraction of the discussed stars, one more periodicity is detected, centered at the sub-harmonic frequency, $\fsh\approx 1/2\fx$. In a model proposed by \cite{dziembowski2016}, the signal detected at $\fsh$ corresponds to a non-radial mode of moderate degree ($\ell=7,\, 8$ or $9$), while the signal at $\fx$ corresponds to its harmonic which typically gains larger amplitude than non-radial mode due to geometrical effects. 

\begin{figure}
\centering
\includegraphics[height=6cm,width=9cm]{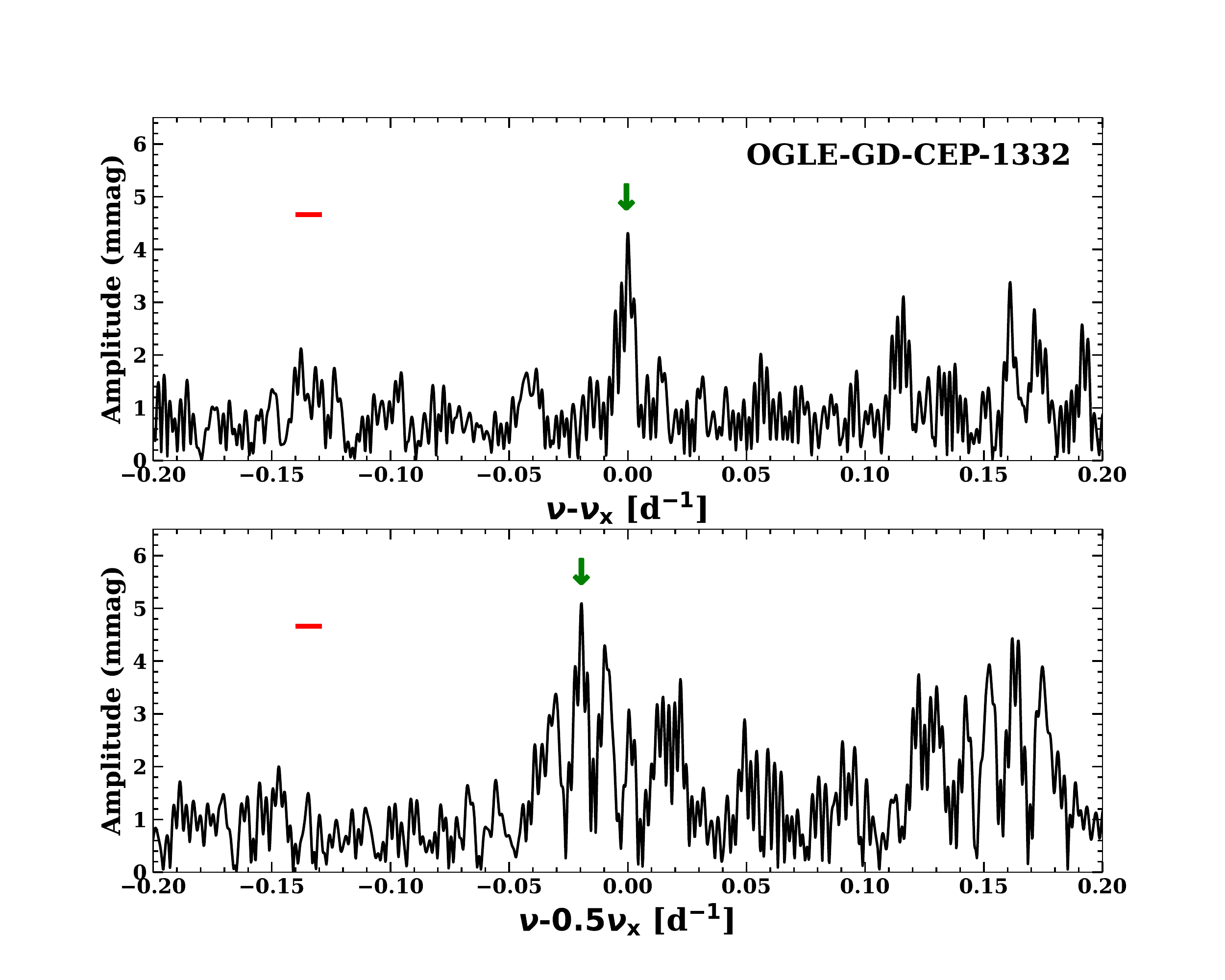}
\includegraphics[height=6cm,width=9cm]{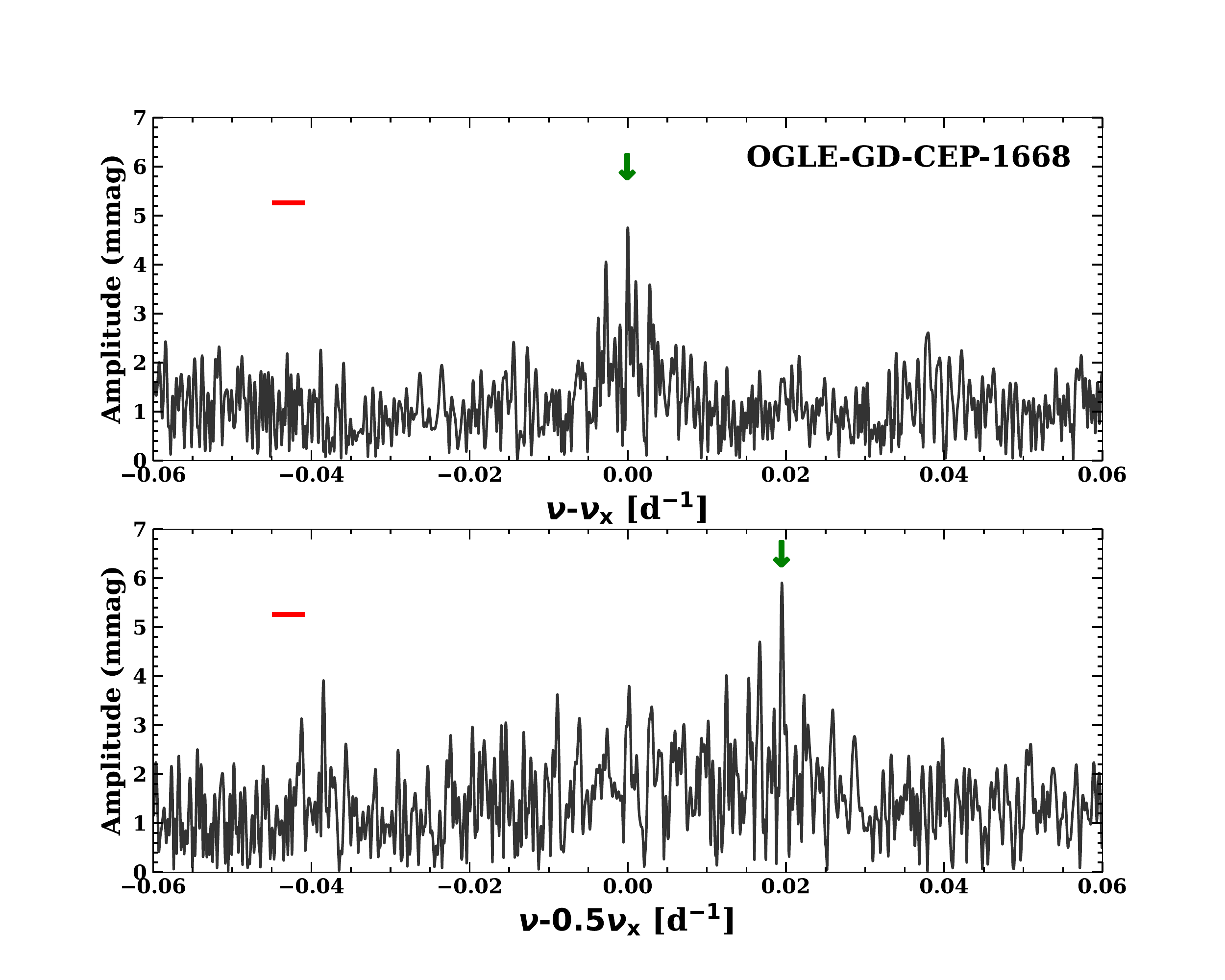}
\caption{Frequency spectra of OGLE-GD-CEP-1332 centred at additional periodicity (first panel) and at its sub-harmonic (second panel). Note that the power excess visible at $\nu-0.5\fx\approx0.15$ in the second panel corresponds to a daily alias of the power excess centered at subharmonic frequency. The green arrow marks the peaks added to the frequency solutions. Red bar location shows the $\sn=4$ level. The third and the fourth panel are the same but for OGLE-GD-CEP-1668.}
\label{fig: OGLE-GD-CEP-1668}
\end{figure}

We detected two candidates, OGLE-GD-CEP-1332 and OGLE-GD-CEP-1668, in which we found simultaneous presence of $\fx$ and $\fsh\approx 1/2\fx$ in the frequency spectra  -- see Fig.~\ref{fig: OGLE-GD-CEP-1668}. The signals detected at $\fx$ place both stars within the middle sequence in the Petersen diagram, see Fig.~\ref{fig: Zoom_non_radial}, where both stars are marked by filled circle enclosed by square, and both period ratios, ie. $\Px/\Po$ (fits the middle sequence) and $\Po/\Psh$ ($\Psh=1/\fsh$) are plotted. Signals detected at $1/2\fx$, in particular in OGLE-GD-CEP-1332 have a complex structure; appear as a power excess, rather than a single and coherent peak, which is also the charactersitics of these signals in Magellanic Clouds Cepheids. Following earlier work of  \cite{smolec2016non} we characterise such signals using the highest peak within the power excess, marked with green arrow in Fig.~\ref{fig: OGLE-GD-CEP-1668}. Even though for OGLE-GD-CEP-1332 we have  $\fsh/\fx=0.4774$, using the frequency of the highest peak for $\fsh$, it is clear that the detected power excess is centred at the subharmonic frequency. For OGLE-GD-CEP-1668, the power excess is not well visible,  but $\fsh/\fx=0.5328$ is still within the range found by \cite{Ziolkowska2020past.conf...75Z}.

Interestingly, there were 9 other candidates with significant detection of additional low-amplitude periodicities, with period ratios falling in a similar area in the Petersen diagram, where $\Po/\Psh$ for OGLE-GD-CEP-1332 and OGLE-GD-CEP-1668 are placed. These stars are marked by `Group 2' in Fig.~\ref{fig: Zoom_non_radial} and the same remark is added in relevant rows in Tab.~\ref{tab: 1O mode data table}. These are firm detections with $\sn>4$; average amplitude exceeds 5 per cent of first overtone amplitude. In these stars, the additional variability usually has a complex structure in the frequency spectrum, just as is typically observed for $\fsh$ signals discussed above. In fact, we speculate that our `Group 2' may represent the same type of variability: we detect a signal at subharmonic frequency \citep[direct detection of non-radial mode according to the model by][]{dziembowski2016}, but the corresponding harmonic has too low amplitude to be detected. This claim is further supported with the comparison to SMC and LMC samples, as analysed by \cite{smolec2016non} and \cite{Ziolkowska2020past.conf...75Z}, respectively. In the Petersen diagram in Fig.~\ref{fig: Zoom_non_radial}, in addition to period ratios in the (0.60,\, 0.65) range, $\Po/\Psh$ period ratios are plotted for stars in which detections at $\fx$ and $\fsh\approx 1/2\fx$ were done simultaneously. Part of our `Group 2' clearly overlaps with SMC/LMC  $\Po/\Psh$ period ratios and part is shifted towards longer periods. This shift seems to follow a metallicity-related trend that is apparent when comparing the SMC and LMC stars: more metal rich LMC stars are shifted towards longer first overtone periods. Galactic Cepheids are on average more metal rich than LMC Cepheids and hence, a shift to even longer first overtone periods.

\subsection{Additional periodicities in F, F+1O and 1O+2O pulsators}
\label{subsec: Additional periodicities in other radial mode sample}

In some stars, after prewhitening with the radial mode(s), additional periodicity is detected at direct vicinity of the prewhitened radial mode frequency. Such detections have period ratio with respect to radial mode period close to unity and are not included in the Petersen diagram in Fig. \ref{fig: 1O+X and subharmonic}. These additional periodicities may arise eg. due to close non-radial mode or due to modulation of the radial mode present. To claim the periodic modulation of pulsation (a few cases are discussed in the next Section), we require at least two additional peaks in the frequency spectrum that would correspond to modulation with the same period. This is not the case for stars discussed here. Examples of such candidates include OGLE-GD-CEP-0106, OGLE-BLG-CEP-009 and OGLE-BLG-CEP-192 in 1O+2O sample; OGLE-BLG-CEP-027 and OGLE-BLG-CEP-196 in 1O sample.

Remaining candidates from F+1O and 1O+2O pulsators sample, which have additional periodicities detected with no radial association, are marked by hollow plus symbols in Fig. \ref{fig: 1O+X and subharmonic}. Their period ratios do not fall directly into the literature radial double-mode sequences. In one F-mode candidate (OGLE-GD-CEP-0079), the radial mode pre-whitened frequency spectra shows an additional variability of longer period. Hence, while period ratio, $\Pf/\Px=0.7233$ places the star within the F+1O double-mode sequence, this interpretation cannot be correct. The other F mode Cepheid shows the additional periodicity located in the `Group 2' region, but the origin of additional periodicity must be different than discussed in the previous Section as the primary pulsation mode is radial fundamental rather than radial first overtone mode.

One candidate (OGLE-GD-CEP-1743) from F+1O sample show additional periodicity nearly in the same location on the Petersen diagram as the `Group 2' region. The interpretation can be the same as discussed in the previous Section, because its period ratio (0.8182), neighboring one of the classified `Group 2' star, is with respect to first overtone mode (\Po $\approx$ 6.752 d). 

Lastly, in 1O+2O sample, additional periodicity in OGLE-GD-CEP-1610 leads to period ratio with respect to radial second overtone placed in the bottom left part of the Petersen diagram in Fig.~\ref{fig: 1O+X and subharmonic} and with respect to radial first overtone beyond the range included in the plot. Another candidate, OGLE-BLG-CEP-004, which is classified as 1O+2O double-mode star, has much more to offer in the radial mode pre-whitened spectrum. There are at least 6 resolved additional periodicities which do not belong any class discussed so far. Three of these frequencies with the highest $\sn$ are reported in Tab.~\ref{tab: 1O+2O mode data table}. The additional frequencies could be due to contamination by nearby variable star(s).

\subsection{New detection of Cepheids with periodic modulation of pulsation}
\label{subsec: Periodic modulation of pulsation}

\begin{table*}
\caption{Characteristics of frequency spectra: frequencies, amplitudes and interpretation for the detected peaks, for new detection of Cepheids with periodic modulation of pulsation.}
\begin{tabular}{lccr}
\hline
\hline
\textbf{OGLE ID (F mode)} & \textbf{Frequency (d$^{-1}$)} & \textbf{Amplitude (mag)} & \textbf{Interpretation} \\
\hline
 OGLE-GD-CEP-1247 & 0.27653018 &	0.2221 & $\fF$  \\
 &0.34196669 &	0.0240 & $\fF+\fm$  \\
 & 0.21109367 &	0.0219 & $\fF-\fm$  \\
 & 0.61849687 &	0.0284 & $2\fF+\fm$  \\
\hline
\hline
\textbf{OGLE ID (F+1O mode)} &{\textbf{Frequency (d$^{-1}$)}} &{\textbf{Amplitude (mag)}} & \textbf{Interpretation} \\
\hline

 OGLE-BLG-CEP-095 & 2.30794775 & 0.0698	 & \fF  \\
 & 3.03732761 &	0.0743 & $\fo$ \\
 & 2.27290653 &	0.0259 & $\fF-\fm$  \\
 & 5.31023414 &	0.0076 & $\fF+\fo-\fm$ \\
 & 4.58085428 &	0.0036 & $2\fF-\fm$  \\
 & 0.76442107 &	0.0031 & $\fo-\fF+\fm$  \\
 & 2.49982528 &	0.0037 & $\nu_{\rm u1}$\\
 & 4.09487519 &	0.0031 & $\nu_{\rm u2}$\\
 & 2.31727992 &	0.0028 & $\nu_{\rm u3}$\\
 \hline
 \hline
\textbf{OGLE ID (1O mode)} &{\textbf{Frequency (d$^{-1}$)}} &{\textbf{Amplitude (mag)}} & \textbf{Interpretation} \\
\hline

 OGLE-BLG-CEP-196 & 3.94388286 & 0.0747	 & $\fo$  \\
 & 3.90851011 & 0.0252 & $\fo-\fm$ \\
 & 7.85239297 &	0.0029 & $2\fo-\fm$ \\
  & 7.81702022 & 0.0043 & $2\fo-2\fm$ \\

 \hline
\end{tabular}
\label{tab: Modulation candidates}
\end{table*}

In this section we present the results for classical Cepheids with periodic modulation in single and double-mode sample. In the frequency spectrum, periodic modulation of pulsation is characterised by equidistant multiplet structures centered at radial mode frequency (and its harmonics). In the ground-based data these are mainly triplets, rarely quintuplets. These multiplets may be highly asymmetric i.e. the side peaks may occur only on one side of radial mode frequency (so called doublets). See \cite{Benko2011MNRAS.417..974B} for detailed analysis explaining the effect of modulation properties on the appearance of the frequency spectra.

\subsubsection{Modulation in fundamental mode Cepheid}

\label{subsubsec: Modulation F mode}

\begin{figure}
\centering
\includegraphics[height=7.0cm,width=8.8cm]{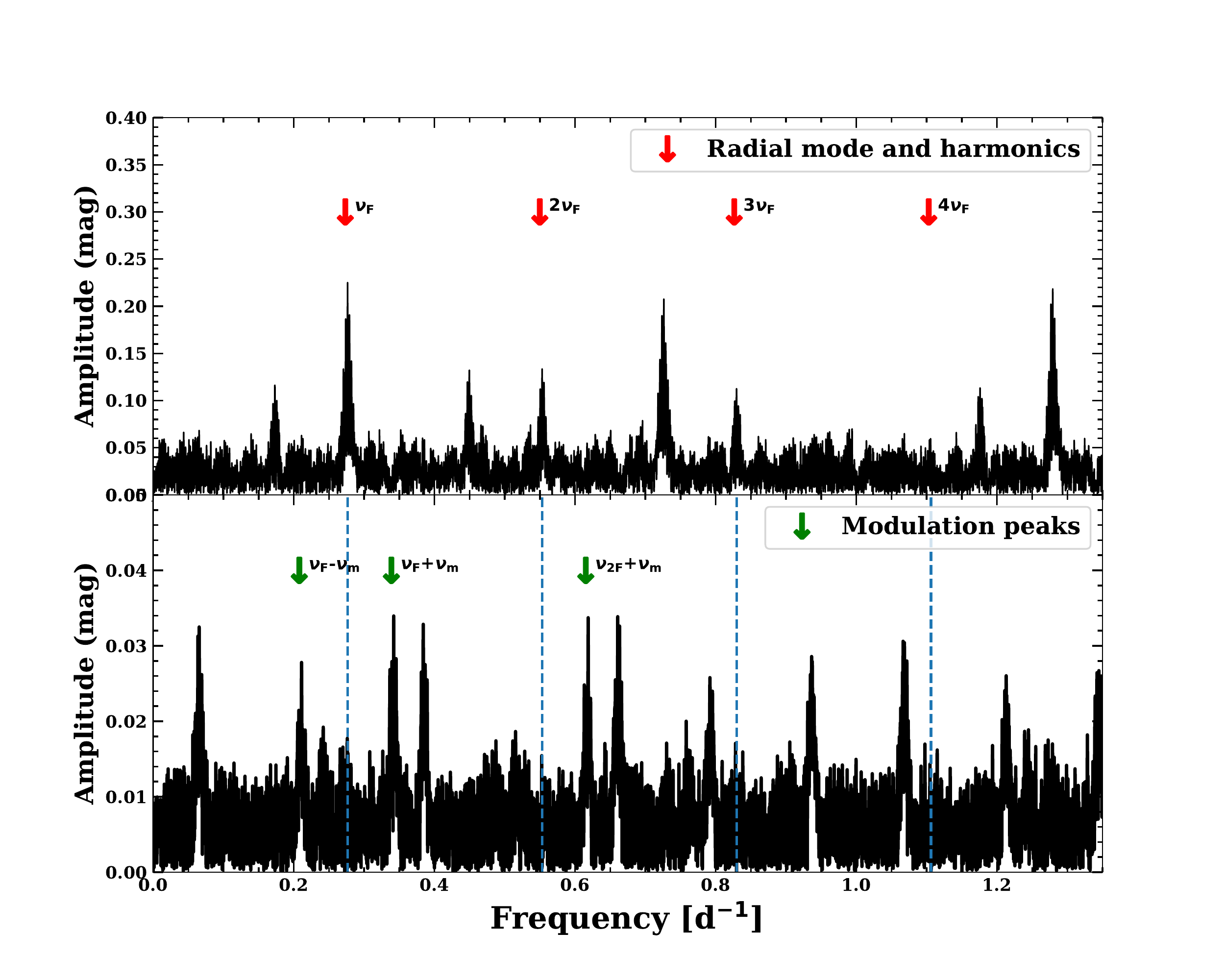}
\caption{Top panel shows radial mode and its harmonics (red arrows) in the frequency spectrum of OGLE-GD-CEP-1247. The bottom panel shows the modulation peaks marked (green arrows) and their spacing with pre-whitened radial mode as well as its harmonics (blue dashed lines).}
\label{fig: OGLE-GD-CEP-1247_freq_spectra}
\end{figure}

\begin{figure}
\centering
\includegraphics[height=6.0cm,width=9cm]{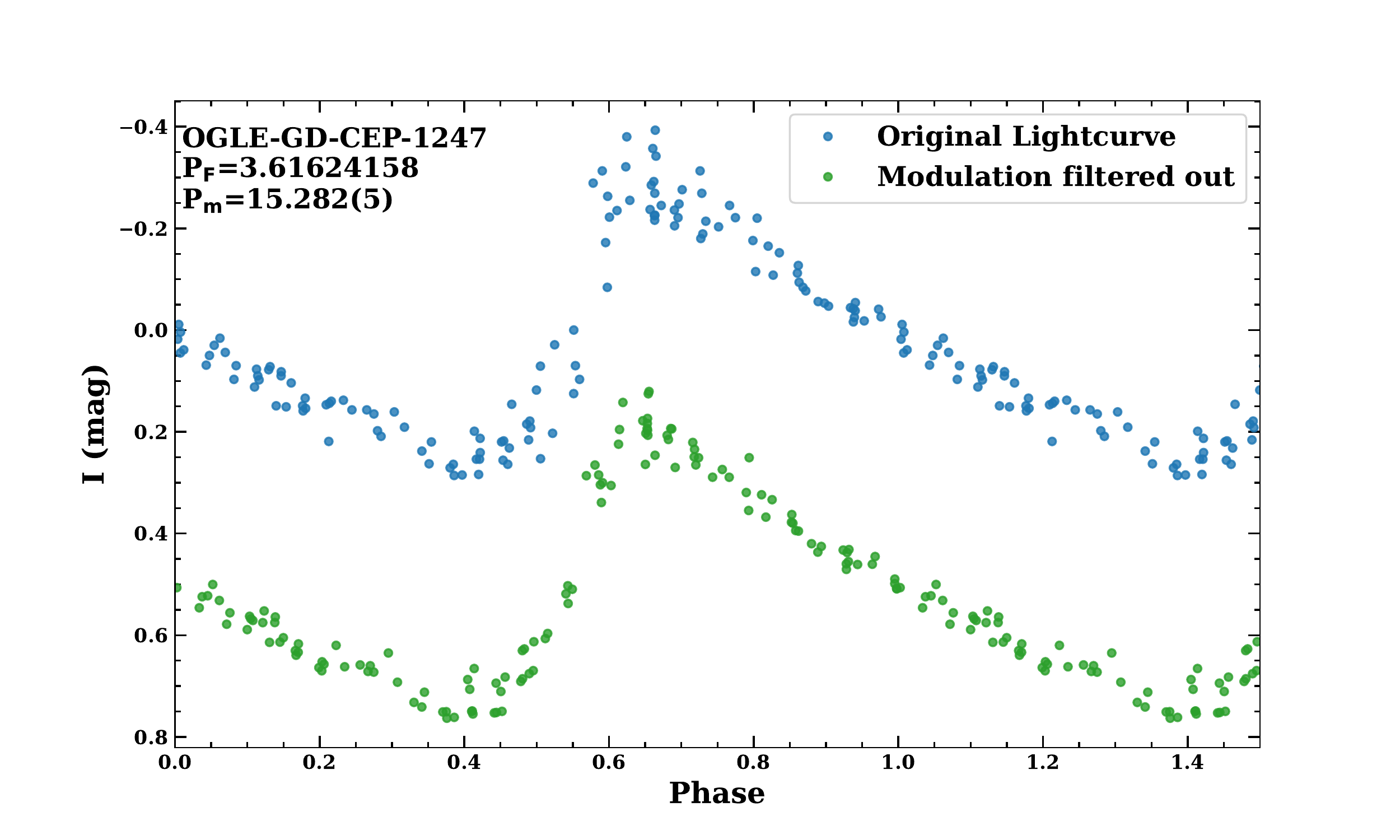}
\caption{ Phased light curve for OGLE-GD-CEP-1247. The light curve in blue denotes the original light curve of the star. The green one is after extracting the modulations from the original light curve.}
\label{fig: OGLE-GD-CEP-1247_modulation_light_curve}
\end{figure}

In the analysis of fundamental mode Cepheids, the sample from Galactic disk revealed an interesting candidate, OGLE-GD-CEP-1247 ($\Pf\approx3.616$\,d), with modulated radial fundamental mode. The modulation ($\Pm=15.282(5)$\,d) can be well seen in the frequency spectra pre-whitened with the fundamental mode and its harmonics (see Fig.~\ref{fig: OGLE-GD-CEP-1247_freq_spectra}). In radial mode pre-whitened frequency spectra, the modulation signature manifest as equidistant triplet around the fundamental mode (peaks at $\fF-\fm$ and $\fF+\fm$) and as a doublet at the first harmonic ($2\fF+\fm$) (data provided in Tab.~\ref{tab: Modulation candidates}). All modulation side peaks are of comparable amplitude. The relative modulation amplitude defined as, $\max(A_-,A_+)/\Af$ (where $A_-$ and $A_+$ are amplitudes of side peaks at $\fF-\fm$ and $\fF+\fm$, respectively) is  10.9\,per cent. After adding these modulation peaks in the frequency solution, we generated the overall light curve along with a modulation signature subtracted one (see Fig. \ref{fig: OGLE-GD-CEP-1247_modulation_light_curve}). Since, it is a low amplitude modulation, the changes in the light curves are marginal, but a lower scatter for light curve with modulation filtered out is noticeable.

\subsubsection{Modulation in double-mode Cepheid} 
\label{subsubsec: Modulation double mode}

\begin{figure}
\centering
\includegraphics[height=4.5cm,width=9cm]{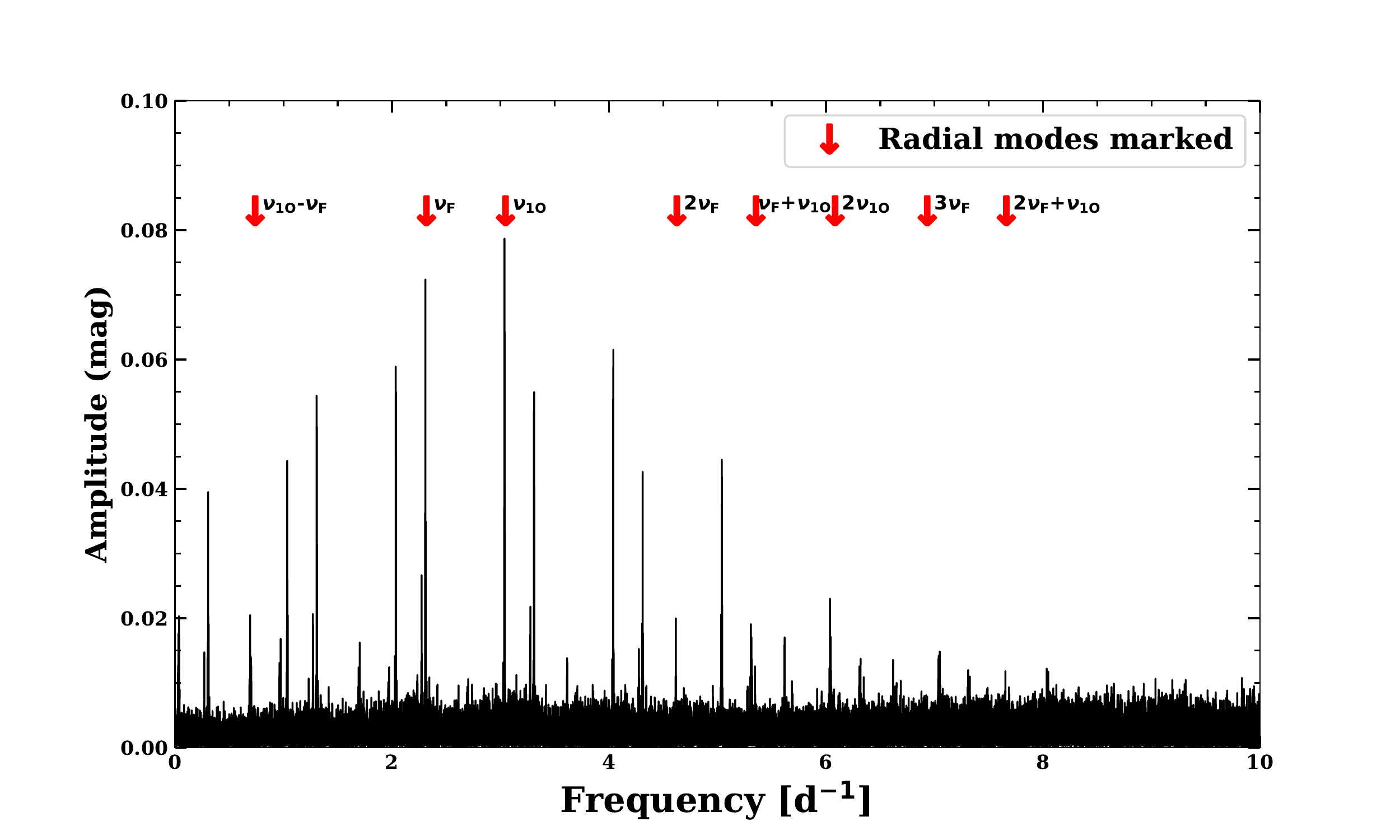}
\includegraphics[height=7cm,width=9cm]{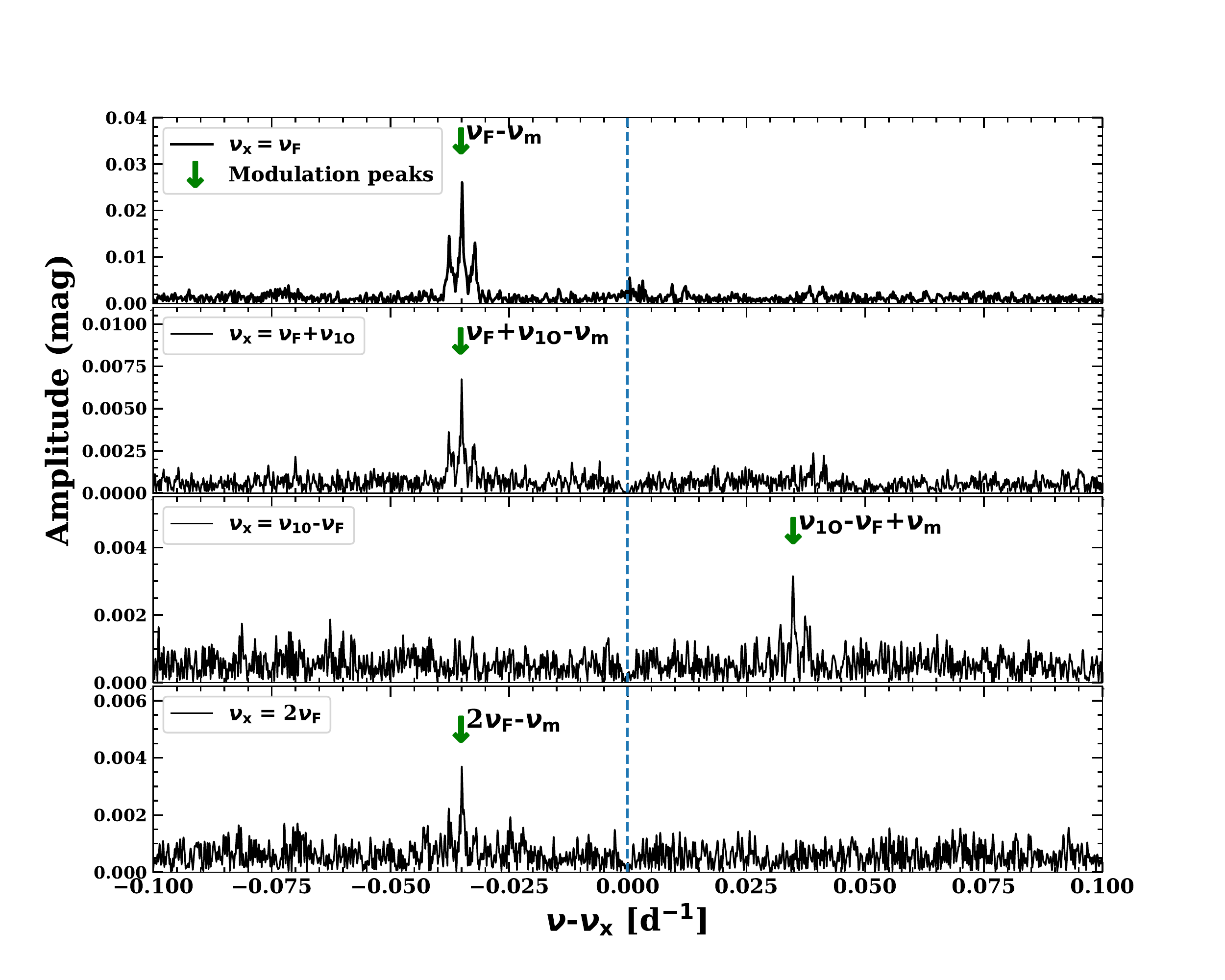}
\caption{Top panel shows radial modes, their harmonics and combination frequencies (red arrows) in the frequency spectrum of OGLE-BLG-CEP-095. Bottom panel shows zoom in of the modulation peaks (green arrows) and their constant spacing with pre-whitened radial mode, its harmonic and combination frequencies (blue dashed lines).}
\label{fig: OGLE-BLG-CEP-095}
\end{figure}

In the double-mode inventory of the Galactic bulge sample we found a modulation candidate OGLE-BLG-CEP-095. This Cepheid pulsates in fundamental and first overtone radial modes with periods $0.433$\,d and $0.329$\,d, respectively. After pre-whitening the frequency spectra with these two modes along with their harmonics and combination frequencies, additional peaks that can be attributed to periodic modulation of the radial fundamental mode with a period of $\Pm=28.5(4)$\,d were detected -- see Fig.~\ref{fig: OGLE-BLG-CEP-095} and Tab.~\ref{tab: Modulation candidates}. Modulation side peaks appear only on lower frequency side of the fundamental mode frequency. Four modulation peaks are detected, including two at combination with radial first overtone frequency: $\fF-\fm$, $2\fF-\fm$, $\fF+\fo-\fm$ and $\fo-\fF+\fm$. The relative modulation amplitude of the fundamental mode, $A_-/\Af$, is 37.1\,per cent. First overtone is not modulated.

After adding these modulation peaks to the solution and pre-whitening, there were still at least three additional signals in the spectra above the detection limit ($\sn>4$) that do not originate due to modulation discussed above ($\nu_{\rm u1}$, $\nu_{\rm u1}$ and $\nu_{\rm u3}$ in Tab.~\ref{tab: Modulation candidates}). Corresponding periods are 0.4000; 0.2442 and 0.4315 days respectively. The period ratios with respect to radial fundamental and radial first overtone period do not fit within the sequences discussed in Section \ref{subsec: Cepheids with double periodic pulsation}. The nature of these signals remains unexplained.

\subsubsection{Modulation in first overtone Cepheid} 
\label{subsubsec: Additional or Modulation}

\begin{figure}
\centering
\includegraphics[height=8.0cm,width=9cm]{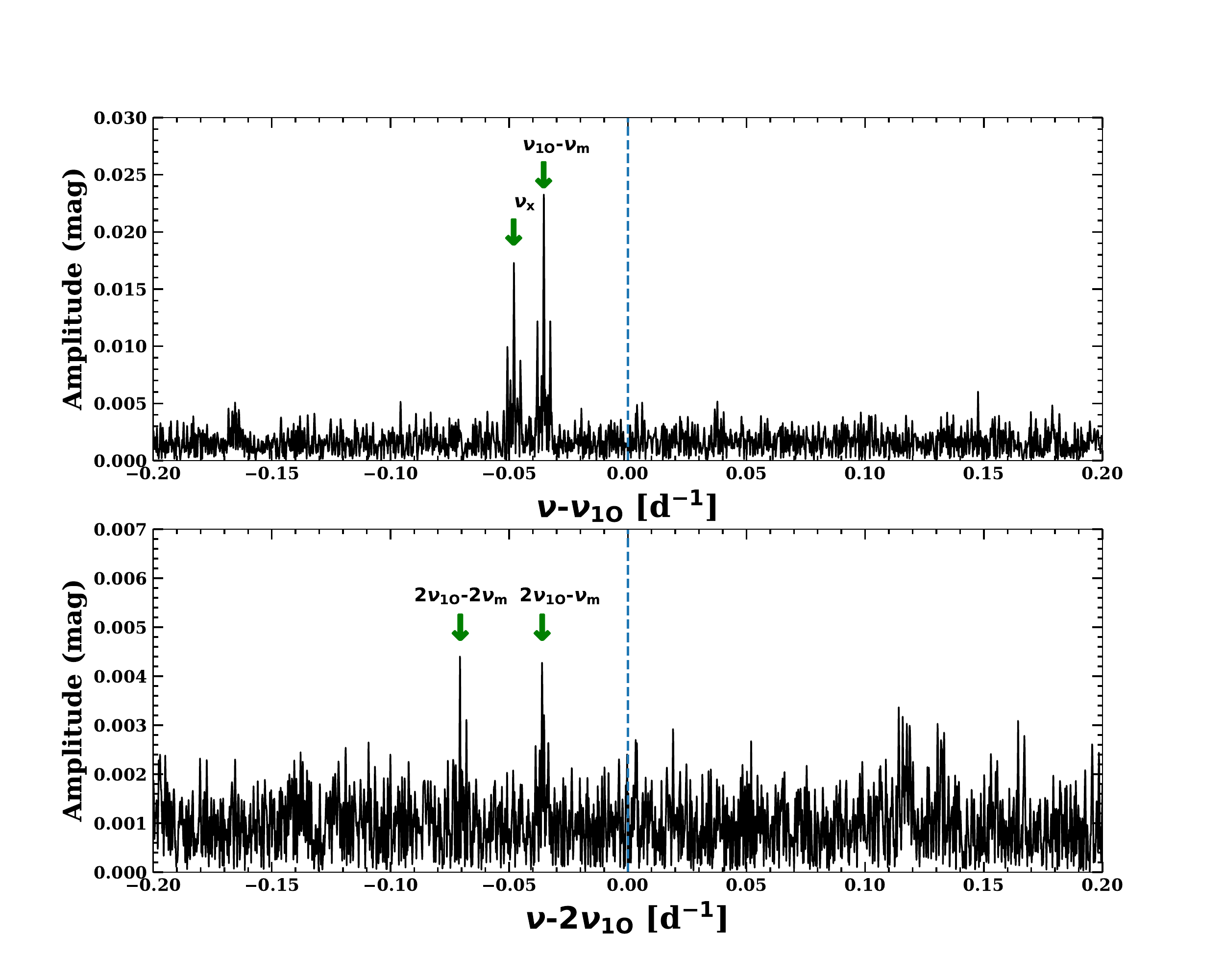}
\caption{Frequency spectrum of first overtone candidate OGLE-BLG-CEP-196 with periodic modulation detection. Top panel shows modulation peak on lower frequency side of the pre-whitened first overtone mode (blue dashed lines) and the peak of unknown origin marked as $\nu_{\rm x}$.  Bottom panel shows the modulation side peaks detected at the harmonic.}
\label{fig: OGLE-BLG-CEP-196}
\end{figure}

In OGLE-BLG-CEP-196, after pre-whitening with the first overtone radial mode  ($\Po=0.254$\,d) and its harmonic, four significant peaks are detected in the frequency spectrum -- see Fig.~\ref{fig: OGLE-BLG-CEP-196}. Three of them can be attributed to a modulation of radial first overtone with a period of $\Pm=28.27(1)$\,d. Their frequencies are $\fo-\fm$, $2\fo-\fm$ and $2\fo-2\fm$. All modulation peaks are located on the lower frequency side of the first overtone frequency. Interestingly, at the harmonic, we detect quintuplet components. Relative modulation amplitude, $A_-/\Ao$, is 33.7\,per cent. The fourth peak ($\sn=12.6$) is located close to $\fo-\fm$ (see upper panel in Fig.~\ref{fig: OGLE-BLG-CEP-196}). It may correspond eg., to non-radial mode, or to secondary modulation, but since we find no combinations with radial mode frequency no firm interpretation is possible.

\section{DISCUSSION}
\label{sec: Discussion}

\subsection{Incidence rate of $\Px/\Po\in (0.60,\, 0.65)$ double-periodic Cepheids in the Galactic field
\label{subsec: Comparison with LMC/SMC}} 

\begin{figure*}
\centering
\includegraphics[height=5cm,width=8cm]{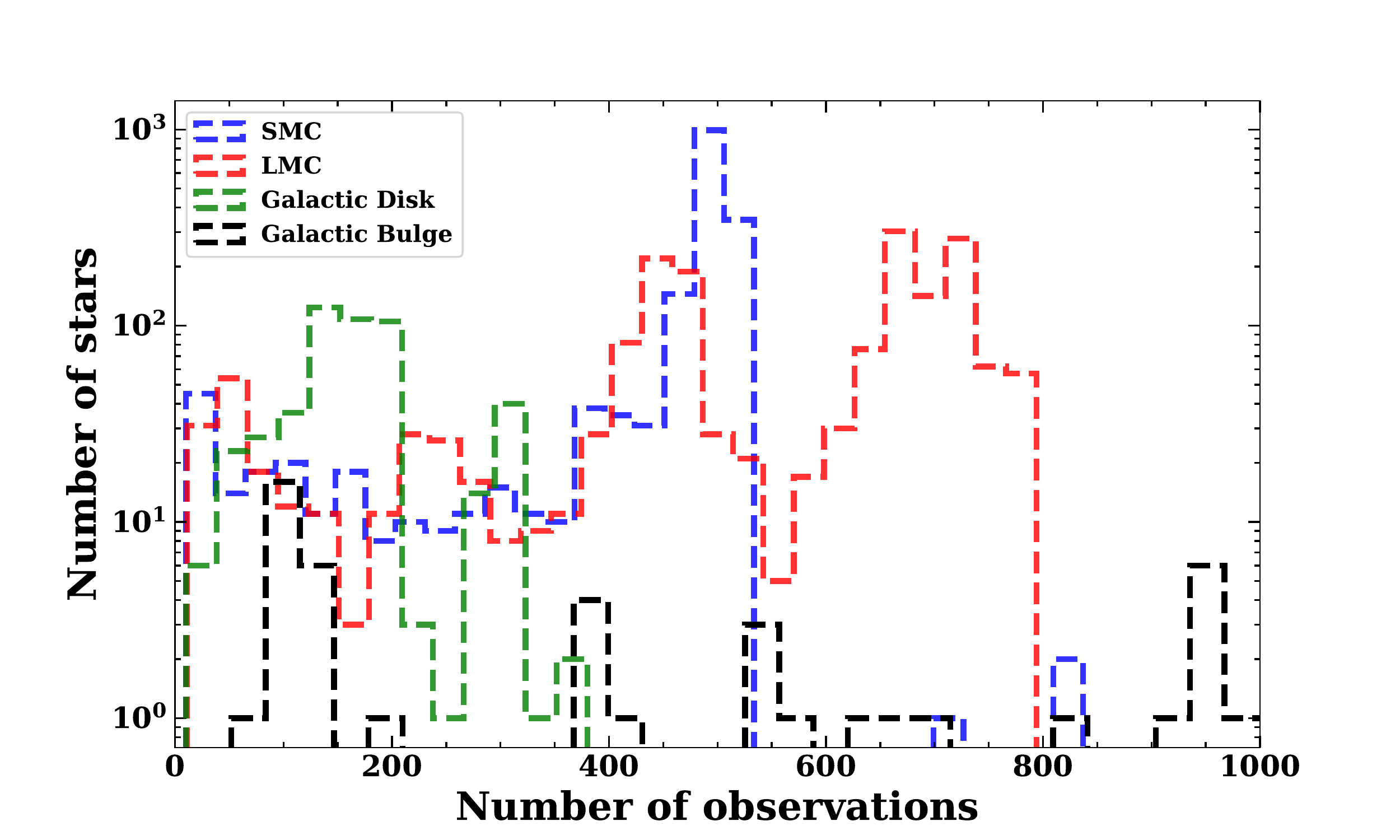}
\includegraphics[height=5cm,width=8cm]{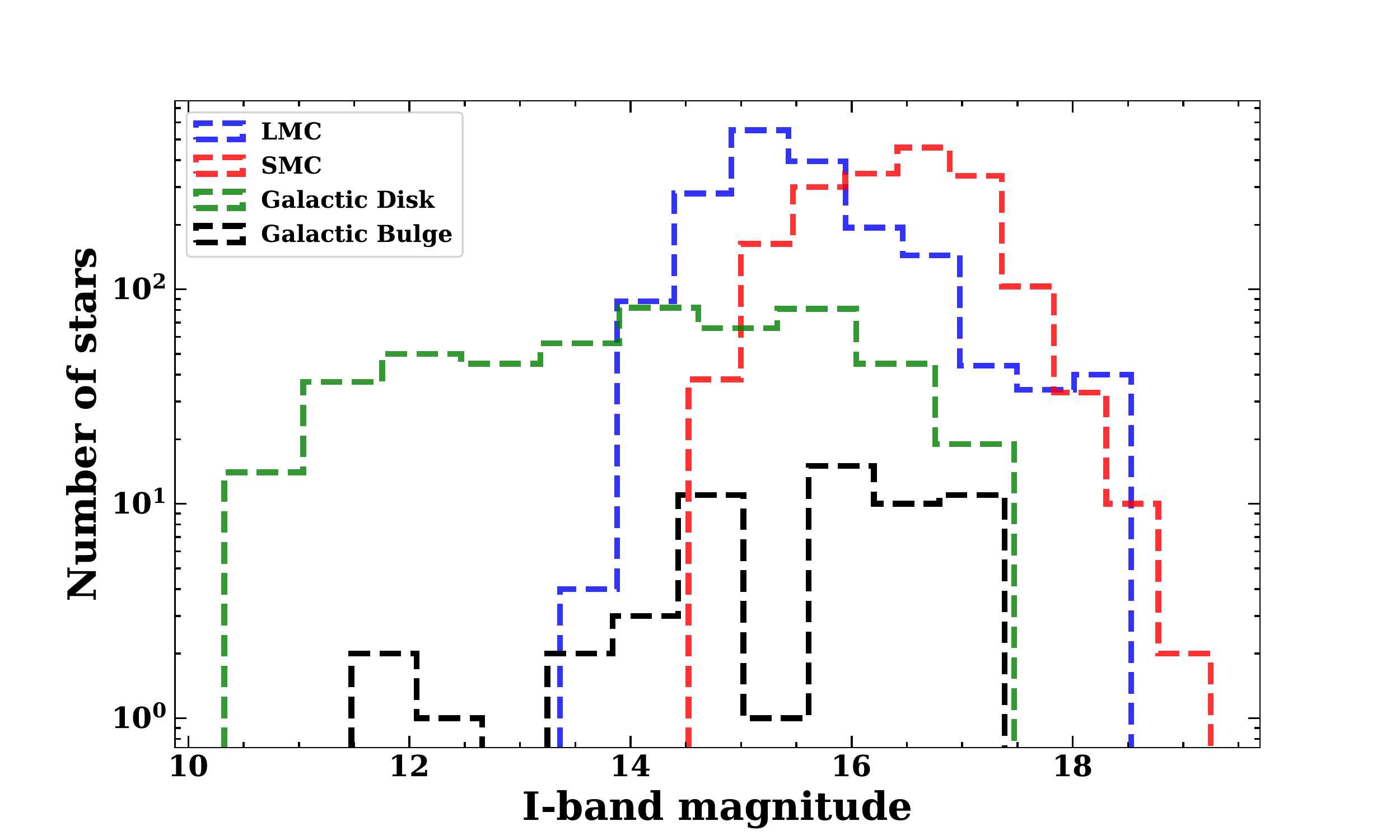}

\caption{Left panel shows histogram of OGLE-IV first overtone Cepheid sample with the number of observations in the SMC (blue), LMC (red), Galactic disk (green) and Galactic bulge (black). Right panel shows the histogram with same colour scheme but for I-band magnitude.} 
\label{fig: Histogram}
\end{figure*}

\begin{figure}
\centering

\includegraphics[height=5cm,width=8cm]{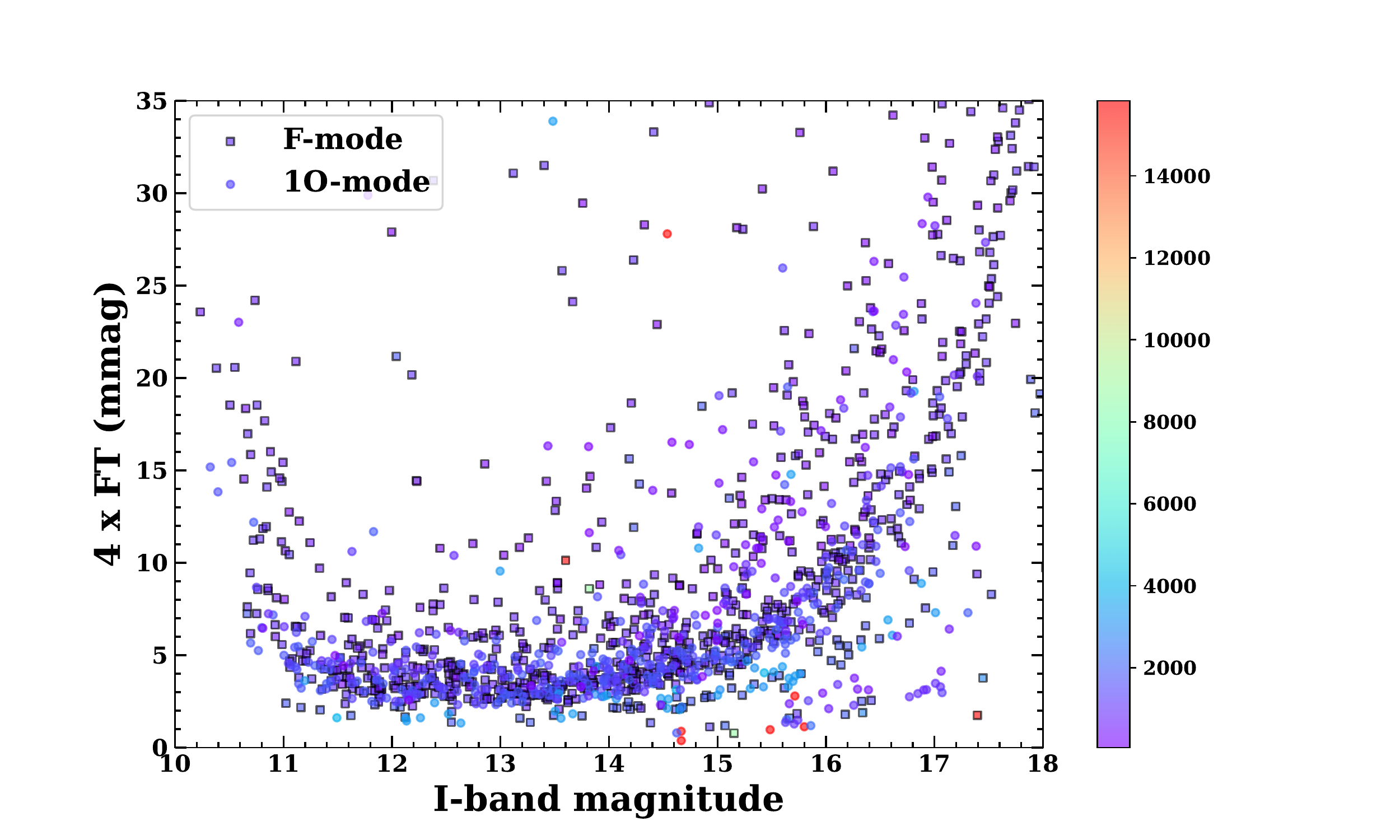}

\caption{The figure represents 4-$\sigma$ noise level computed as a function of mean I-band magnitude for Galactic Cepheids pulsating in fundamental (filled square) and first overtone (filled circle) modes. The colour bar represents number of observations for each star.} 
\label{fig: detection-limit}
\end{figure}

Till now, nearly all double-periodic Cepheids with $\Px/\Po\in (0.60,\, 0.65)$  were discovered in the Magellanic Clouds. \cite{soszynski2010optical} analyzed the OGLE-III data for SMC sample, and detected such stars with an incidence rate of 8.4\,per cent (in 138 out of 1644 1O Cepheids). Later, \cite{Suveges2018b} studied Magellanic Clouds using local kernel regression pre-whitening method \citep{Suveges2018a}. Considering their unambiguous detections only, incidence rates are 6.9\,per cent (120 stars out 1739) in the LMC and 7.8\,per cent (139 stars out of 1782) in the SMC. The most recent, but incomplete investigation of the above mentioned additional variability in Cepehids using SMC/LMC OGLE-IV data by \cite{Ziolkowska2020past.conf...75Z} reported incidence rates of 21.0\,per cent (205 out of  976 analysed stars) for LMC and 11.0\,per cent (168  out of 1522 analysed stars) for SMC. \\

The incidence rates given above can only serve as lower limits. \cite{soszynski2008optical} did not intend to comprehensively search for these additional periodicities. The incidence rates from \cite{Ziolkowska2020past.conf...75Z} are derived from preliminary and incomplete analysis. Moreover, analysis by \cite{Suveges2018b} only focused on characterising and interpreting only a single additional peak beyond the radial mode, hence other additional signals may be missed and the numbers not providing the accurate incidence rate of the complete sample.

Now bringing the discussion to Galactic Cepheids, let us first note that we have detected double-periodic Cepheids of the discussed class in the Galactic disk fields only (no detection in the Galactic bulge). The overall Galactic disk Cepheid sample is order of magnitude smaller than the Magellanic Cloud samples. For context, total Galactic disk 1O sample is roughly 2.6 times smaller than their LMC/SMC counterparts. Incidence rates in Galactic disk sample turns out to be 2.4\,per cent (12 stars out of 495). The incidence rate is thus much smaller than in the Magellanic Clouds. To get insight into reason behind significantly smaller incidence rate we need to look at the detection limit and compare it with the ones in Magellanic Clouds.\\

There are essentially two factors that affect the noise level in the frequency spectrum of a Cepheid: number of observations and mean brightness. The noise level scales with $N^{-1/2}$, where $N$ is the total number of observations for each Cepheid. To to put this into context, Fig. \ref{fig: Histogram} left panel shows the histogram of number of observations per Cepheid, in OGLE-IV survey, based on observing fields. The histogram is plotted for 1O Cepheids in SMC (blue), LMC (red), Galactic disk (green) and Galactic bulge (black). We can see that there are larger number of observations in SMC/LMC as compared to Galactic disk and bulge fields. Hence, for Cepheids of similar brightness, one can expect that on an average the mean noise level in the frequency spectra of Galactic Cepheids will be higher than in the SMC/LMC. The right panel of Fig. \ref{fig: Histogram} depicts the mean I-band brightness distribution of Cepheids from the above mentioned four observing fields. Cepheids in the Galactic disk cover a much larger range of I-band magnitudes, including bright targets. At similar number of observations, the brighter the Cepheid, the lower the noise level is to be expected.

For direct estimation of the detection limit, we follow the approach given in \cite{smolec2016non} (see their fig. 13) and use 1O Cepheid sample from the Galactic field to do a homogeneous analysis where we fit 6$^{\text{th}}$ order Fourier series of the corresponding principle frequency (i.e. \fo). We also remove possible trends in the data along with first overtone associated non-stationary signals using time-dependent pre-whitening on a season-to-season basis. Finally, we compute the frequency spectrum of the pre-whitened data in the (0,\,3\fo) range and record the average noise level which, multiplied by 4, corresponds to our detection limit. In Fig.~\ref{fig: detection-limit}, it is represented as a function of both mean I-band magnitude and number of observations. The estimation of the detection limit for fundamental mode stars follows a similar procedure with difference in fitting a 14$^{\text{th}}$ order Fourier series (to account for more non-linear light curve shape) as performed in  \cite{smolec2017unstable} (see their fig. 5). With this exercise, we can clearly see that for fainter stars, the noise level increases. Whereas, at the same brightness, the noise level of some bulge Cepheids is low, since they have higher number of observations.
Qualitatively, if we compare our detection limit for 1O sample with \cite{smolec2016non} (their fig. 13) and for F-mode sample with \cite{smolec2017unstable} (their fig.~5), we conclude that the detection limit for Galactic sample is, on average, significantly higher and hence the chances to detect additional low-amplitude periodicities are lower in the Galactic fields. Consequently, the reported incidence rates are also significantly smaller.\\

We note that period ratios in the 0.60--0.65 range exist in first overtone RR Lyrae stars as well, known as RR$_{0.61}$ group. \cite{netzel2019census} quantified their occurrence to a total of $\sim$8.3\,per cent (960 stars out of 11,415) for the Galactic bulge OGLE-IV sample. This fraction is a lower limit. For high cadence fields, the detection limit is much lower and this fraction is as high as 27\,per cent \citep{Netzel2015bMNRAS.453.2022N}. Even higher incidence rates of 100 per cent, are found for space observations \citep{Molnar2015MNRAS.452.4283M,moskalik2015kepler}.

\subsection{Galactic Cepheids with non-radial modes}
\label{subsec: Comparison of non-radial mode }

Till now, more than 300 Cepheids pulsating in 1O mode have been detected to be exhibiting additional periodicity with period ratio $\Px/\Po\in (0.60,\, 0.65)$. These stars were nearly exclusively detected in the Magellanic Clouds \citep{soszynski2008optical,soszynski2010optical,soszynski2016multi,moskalik2009frequency,smolec2016non,Suveges2018b,Ziolkowska2020past.conf...75Z}. Our analysis reports only 12 such candidates in the Galactic field. Nevertheless, our detections significantly increases the sample of Galactic 1O Cepheids with period ratio of 0.60--0.65, as prior to this, there was only one detection reported by \cite{Pietrukowicz2013AcA....63..379P}. \\

The discussed stars form three sequences in the Petersen diagram. \cite{Suveges2018b} reported a systematic shift (see their fig. 8) in all three sequences when SMC and LMC candidates are compared. The higher metallicity LMC stars are shifted towards longer pulsation periods. As we compare our Galactic sample with that of the OGLE Magellanic Clouds (see Fig. \ref{fig: Zoom_non_radial}), we reinforce the existence of this metallicity-related trend. The sequences from Galactic Cepheids (even more metal-rich, on average, than LMC Cepheids) are even more shifted to higher 1O periods. Hence the picture becomes even more comprehensive. We note that these shifts follow the shift observed for period distribution of all Cepheids in the Magellanic Clouds, see eg. fig. 9 from \cite{soszynski2010optical}.

In the discussed group, additional low-amplitude periodicities (at \fx) are often accompanied with signals centered at sub-harmonic frequency (1/2\fx). The signals at sub-harmonic frequency are almost never found as a coherent peak but rather a group of peaks or a power excess. It indicates that associated variability may be prone to phase and/or amplitude variations. \cite{smolec2016non} reported these sub-harmonic cases to be 35\,per cent in total. In our sample we have two candidates (OGLE-GD-CEP-1332 and OGLE-GD-CEP-1668) with simultaneous presence of both $\fx$ and $1/2\fx$ signals. According to \cite{dziembowski2016} model, the signals detected at sub-harmonic frequencies (1/2\fx) are due to non-radial modes of moderate degrees, $\ell=7,\,8$ and $9$. Direct detection of such non-radial modes is difficult as their observed amplitude is reduced due to geometric cancellation.  It is easier to detect the harmonics, ie. signals at $\fx$, that form three sequences in the Petersen diagram with period ratios in the 0.60--0.65 range. The top sequence arises due to harmonics of $\ell=7$ modes, the middle sequence due to harmonics of $\ell=8$ modes and the bottom sequence due to harmonics of $\ell=9$ modes. Geometric cancellation is the lowest for even-$\ell$ modes, so it should be relatively easy to detect sub-harmonics, ie., to directly detect non-radial modes, for Cepheids of the middle sequence. It is indeed the case as analysed by \cite{smolec2016non} who reported 74 per cent stars of the middle sequence accompanied with simultaneous presence of $1/2\fx$ signal. For the top and bottom sequences the corresponding numbers are 31 and 8 per cent, respectively, again in agreement with \cite{dziembowski2016} model. Interestingly, the two cases we report here correspond to the middle sequence. While the statistics is low, it well agrees with predictions of the model.

As we already argued earlier, stars of our `Group 2`, most likely correspond to direct detection of non-radial modes. In these stars harmonics, are not detected, as likely their amplitudes fall below our detection limit. It is interesting to note that their numbers are quite comparable to the overall number of detections in `Group 1', ie. double-periodic Cepheids with period ratios $\Px/\Po\in (0.60,\, 0.65)$. The amplitude comparison analysis ($A_{\rm sh}$ vs. $A_{\rm x}$) by \cite{smolec2016non} reported cases where the sub-harmonic signal amplitude is larger than its harmonic (see their fig. 12). \cite{Suveges2018b} also reported a prominent group of double-periodic Cepheids, called 1.25-modes, that seem equivalent to our `Group 2' and discussed their connection to 0.62-modes, i.e. double-periodic stars with period ratios in the 0.60--0.65 range. Considering the placement of sub-harmonic signals detected in the Magellanic Clouds, and of our `Group 2' in the Petersen diagram (Fig. \ref{fig: Zoom_non_radial}), we note a shift of our `Group2' towards longer pulsation periods. It well agrees with the metallicity-related shift  for the linked three sequences with 0.60--0.65 period ratios. An interpretation in which additional periodicities detected in `Group 2' correspond to non-radial modes of $\ell=7-9$ seems most likely.

To complete the picture with RR Lyrae stars, the sub-harmonic detection from Galactic bulge sample constitutes to 11.9\,per cent \citep[114 stars out of 960 RR$_{0.61}$;][]{netzel2019census}, with sub-harmonic detections corresponding to the top sequence being the most populated. It well agrees with the \cite{dziembowski2016} model, in which the top sequence for RR Lyrae stars arises due the harmonic of an even order, $\ell=8$ mode. Interestingly, \cite{netzel2019census} also reported additional signals in RRc stars that can be interpreted as due to direct detection of non-radial mode, without harmonic signal present -- an analogue of our `Group 2'.

\subsection{Galactic Cepheids with modulation}

\label{subsec: Comparison of modulated Cepheids}
The  quasi-periodic modulation in amplitude and/or phase in RR Lyrae stars, known as the Blazhko effect \citep{blazko1907mitteilung} was a motivation to search for similar behaviour in classical Cepheids. Despite recent progress on both observational and theoretical side, see the reviews by \cite{Kovacs2016CoKon.105...61K} and \cite{Smolec2016bpas..conf...22S} and summary of the most promising theoretical model by \cite{Kollath2021}, a fully satisfactory explanation of the Blazhko mechanism is still missing. Till recently, modulations were nearly exclusively associated with RR Lyrae stars. \cite{moskalik2009frequency} discovered periodic modulation of pulsation in double-mode, 1O+2O Cepheids. As both radial modes are modulated with the same period, and their amplitudes are anti-correlated, they linked the origin of these modulations to the resonant interaction either between the radial modes or one radial mode with a non-radial mode. Modulations were also discovered in single-mode 1O Cepheids \citep{soszynski2016multi} and in the second overtone oddball, V473 Lyrae \citep{Molnar2014MNRAS.442.3222M}. Low-amplitude periodic modulation of pulsation was discovered in {\it Kepler} observations of single-mode F-mode Cepheid, V1154 Cyg \citep{Derekas2012MNRAS.425.1312D,Derekas2017MNRAS.464.1553D}, followed by the discovery of a numerous sample of 51 modulated F-mode Cepheids in the OGLE sample of Magellanic Cloud Cepheids \citep[29 stars in the SMC and 22 stars in the LMC;][]{smolec2017unstable}. In the Magellanic Cloud sample, modulations are very frequent in stars with pulsation periods around 10 days. Based on that, \cite{smolec2017unstable} suggested that the modulations may be caused by the. 2:1 resonance between the fundamental mode and the second overtone, which is also responsible for Cepheid bump progression \citep{simon1976ApJ...205..162S, buchler1986ApJ...303..749B, Buchler1990ApJ}.

\begin{figure}
\centering
\includegraphics[height=7cm,width=9cm]{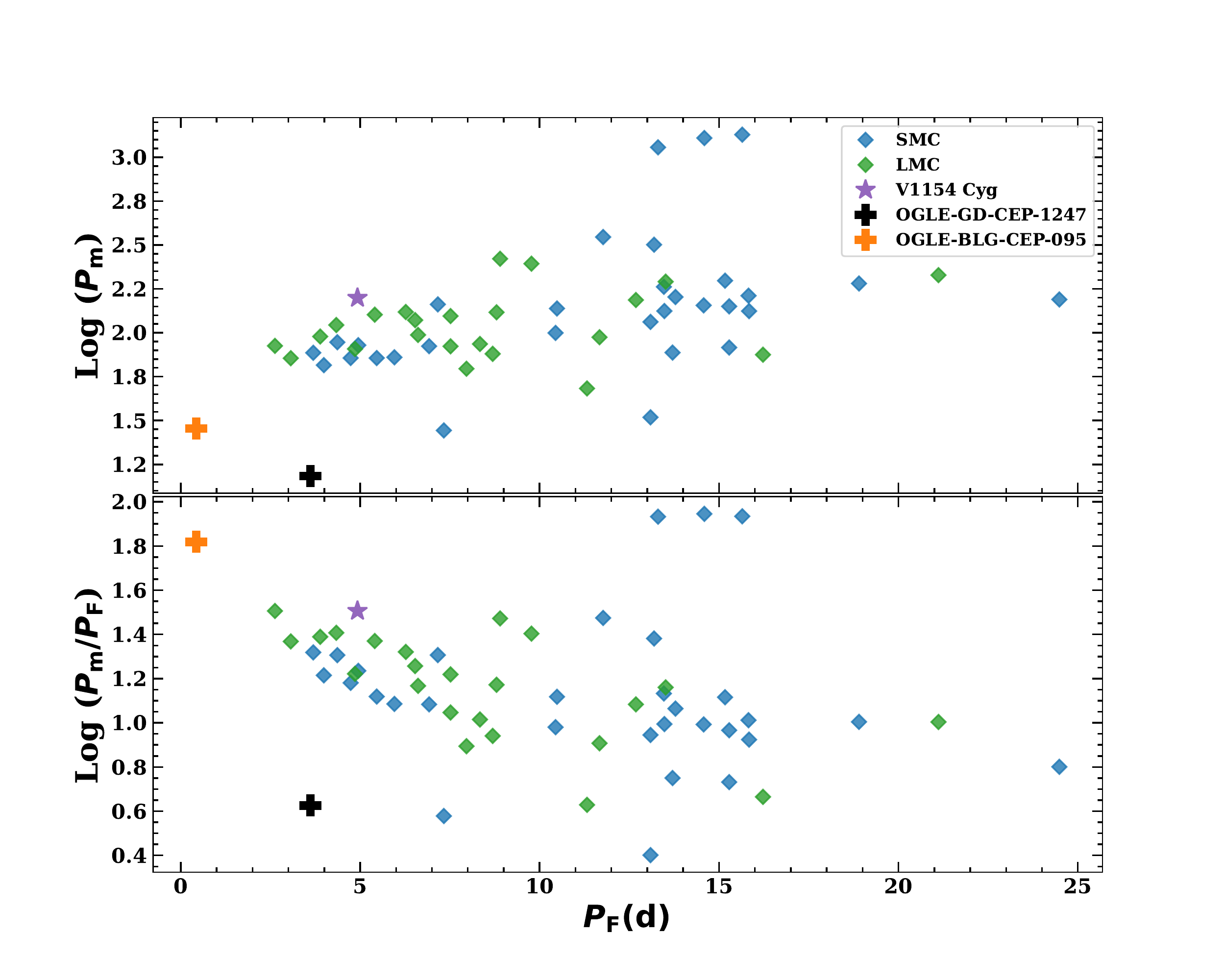}
\caption{Upper panel shows comparison of modulation period as a function of fundamental mode period in Cepheids in SMC (blue diamonds) and LMC (green diamonds) from \citet{smolec2017unstable}, with OGLE-GD-CEP-1247 (black plus) and OGLE-BLG-CEP-095 (orange plus). V1154 Cyg, a \textit{Kepler} detection is marked with purple star symbol. Lower panel has the same colour scheme but for ratio of modulation period to F-mode pulsation period.}
\label{fig: Modulation literature comparison}
\end{figure}

We have detected periodic modulation of the fundamental mode in two Cepheids, one pulsating in the F-mode (OGLE-GD-CEP-1247) and one being a double-mode, F+1O Cepheid (OGLE-BLG-CEP-095). The latter detection, is the first of its kind -- no modulated double-mode F+1O Cepheids was known before. Since in both stars it is the fundamental mode that is modulated, we can compare their properties, pulsation and modulation periods, with the Magellanic Clouds sample from \cite{smolec2017unstable}. V1154 Cyg, in which F-mode is modulated as well, is also added to the comparison. As seen from Fig. \ref{fig: Modulation literature comparison}, the fundamental mode period and modulation period of OGLE-BLG-CEP-095 follow the extended trend of the SMC/LMC modulated stars. The \textit{Kepler} detection, V1154 Cyg, also falls along the same progression. It may indicate that the mechanism behind the modulation of pulsation is the same in all discussed Cepheids. However OGLE-GD-CEP-1247, shows a larger departure from the discussed progression. Admittedly, we have only a handful of detections at the moment and no elaborated model behind. 

Considering single-mode fundamental mode Cepheids we have detected only one modulated star, while in the Magellanic Clouds there are many more detections. In particular, \cite{smolec2017unstable} noticed that in the SMC the phenomenon is very common (incidence rate of 40 per cent) in the period range of $12\,{\rm d}<\Pf<16\,{\rm d}$. For higher metallicty LMC, the the highest incidence rate was only 5 per cent for $8\,{\rm d}<\Pf<14\,{\rm d}$. To investigate a possible role of metallicity, we recalculate the incidence rates for a period interval centered on 10 days, $6\,{\rm d}<\Pf<14\,{\rm d}$ to arrive at 6.5 per cent and 4 per cent in the SMC and LMC, respectively. The incidence rate decreases for more metal rich LMC. Assuming the same incidence rate for the Galactic fields as for the LMC, we should detect about 9 modulated Cepheids in the disk and one modulated Cepheid in the bulge (the bulge sample is scarce). We have detected only a single modulated F-mode Cepheid in the disk, OGLE-GD-CEP-1247. One cannot conclude however, that for even more metal rich Galactic disk the incidence rate of modulated Cepheids must be intrinsically lower. This is because of a significantly higher detection limit of Galactic samples as discussed already in Sect.~\ref{subsec: Comparison with LMC/SMC}. The typical amplitude of the modulation peaks in the LMC sample is 2--3 mmag \citep[see tab. 2 in][]{smolec2017unstable}, while for majority of stars in our sample the detection limit is at larger magnitudes, see Fig.~\ref{fig: detection-limit}.

The third detection of modulated Cepheid from our sample, is a single-mode first overtone Cepheid. Several other modulated first overtone Cepheids were reported earlier by \cite{soszynski2016multi}. The main difference between ours and previous detections is in modulation amplitude -- it is small in our detection, and much larger in previous detections, in which modulation is easily visible in lightcurves alone.

\section{SUMMARY AND CONCLUSIONS}
\label{sec: Summary and Conclusions}

The major findings of the analysis from the latest OGLE-IV survey photometry data for Cepheids in Galactic disk and bulge fields is summarized as follows:
\begin{enumerate}
\item Twelve new candidates for double-mode radial pulsation were identified. These include F+1O, 1O+2O and F+2O pulsations.

\item Six triple-mode radial candidates were identified: five F+1O+2O and one 1O+2O+3O. These triple-mode candidates are very rare and yet very valuable assets as they are excellent laboratories for asteroseismology to test stellar evolution theory \citep[eg.][]{MoskalikDziembowski2005A&A...434.1077M}.

\item Admittedly, additional periodicities that we associate with radial pulsations are of low amplitude and no linear combination frequencies with dominant radial frequency are detected. In some cases of double-mode pulsation alternative interpretation, mostly in terms of other radial mode, is possibly due to daily alias ambiguity  (eg. either 1O+2O or 1O+3O pulsation is possible, depending which alias is selected). Consequently, the new candidates should be considered with caution.

\item In twelve first overtone Cepheids we detect additional low-amplitude periodicity of shorter period, corresponding period ratios, $\Px/\Po$, fall in the $(0.60,\, 0.65)$ range. Such double-periodic Cepheids are well known in the Magellanic Clouds and form three sequences in the Petersen diagram.  In the Galactic fields only one such double-periodic Cepheid was known -- our detections significantly increases the Galactic sample. Comparing with Magellanic Cloud Cepheids we note that the more metal rich field, the more shifted are double-periodic stars towards longer periods. According to a model proposed by  \cite{dziembowski2016}, these additional periodicities are harmonics of non-radial modes of moderate degrees, $\ell=7,\, 8$ and 9, which are easier to detect than non-radial modes themselves, as amplitudes of the latter are reduced by geometric cancellation. 

\item In two double-periodic stars with $\Px/\Po\in (0.60,\, 0.65)$ we detect significant signal at sub-harmonic frequency, ie., et $1/2\fx$. According to the model proposed by \cite{dziembowski2016} these signals are direct detections of non-radial modes of moderate degrees. Moreover, in nine first overtone Cepheids we detect additional variability that most likely corresponds to direct detection of the above mentioned non-radial modes, with no signature of signal at the harmonic. 

\item We report the discovery of three Cepheids with low-amplitude, periodic modulation of pulsation. Modulation was detected in single-mode F Cepheid OGLE-GD-CEP-1247 ($\Pf$=3.61624(5)\,d  and $\Pm$=15.282(5)\,d), double-mode F+1O Cepheid, OGLE-BLG-CEP-095 ($\Pf$=0.43329(2)\,d, $\Po$=0.32924(3)\,d and $\Pm$=28.5(4)\,d) and in single-mode, 1O Cepheid, OGLE-BLG-CEP-196 ($\Po$=0.2535572(1)\,d and $\Pm$=28.27(1)\,d). With our sample, the number of known modulated Cepheids in the Galactic fields is significantly increased.  Modulation in the F+1O double-mode Cepheid is the very first detection of modulation in such combination of double-mode Cepheid pulsation. Of the two radial modes, we detect modulation of fundamental mode only.

The homogeneous OGLE photometry for classical Cepheids in the Magellanic Clouds and in the Galactic bulge and disk fields in principle should allow to systematically study how the incidence rate of various phenomena, like excitation of additional non-radial modes or of periodic modulation of pulsation, depends on metallicity. Unfortunately, incidence rates in the Galactic fields turn to be much lower than in the Magellanic fields and do not allow detailed statistical studies of the detected phenomena. This however is not due to intrinsically lower incidence rates in the more metal-rich Galactic fields, but due to significantly larger detection limit. The main reason is lower sampling rates in the Galactic fields and shorter baseline of observations. Consequently, continuous monitoring of Galactic fields, possibly with higher sampling rate is of extreme value to get more insight into still poorly known phenomena of non-radial mode excitation and periodic modulation of pulsation in classical Cepheids.

Since the discussed phenomena are of low amplitude, new detections are also expected from space telescopes, eg., from the ongoing NASA \textit{Transiting Exoplanet Survey Satellite} \citep{Ricker2015JATIS...1a4003R} mission delivering precise photometry of Galactic fields. The mission will enable to closely monitor cycle-to-cycle variations for bright Cepheids. The Cepheid first light results of the TESS mission \citep{Plachy2021ApJS..253...11P} shows promising signs as it detects one candidate ($\beta$ Dor), with cycle-to-cycle variations of the F-mode. An anomalous Cepheid (XZ Cet) pulsating in first overtone mode was also detected to contain low amplitude non-radial mode of 0.61-type.

\end{enumerate}

\section*{Data availability}
The photometry data for OGLE-IV Galactic Cepheids \citep{udalski2015ogle,soszynski2017AcA....67..297S,Udalski2018AcA....68..315U,soszynski2020AcA....70..101S} used in this paper is publicly available and can be downloaded from \url{ftp://ftp.astrouw.edu.pl/ogle/ogle4}. Other data products will be shared on reasonable request to the corresponding author.

\section*{Acknowledgements}
This research is supported by the National Science Center, Poland, Sonata BIS project 2018/30/E/ST9/00598. H.N. is supported by the Polish Ministry of Science and Higher Education under grant 0192/DIA/2016/45 within the Diamond Grant Programme and by the Foundation for Polish Science (FNP).



\bibliographystyle{mnras}
\bibliography{mnras} 





\bsp	
\label{lastpage}
\end{document}